\documentclass[fleqn,usenatbib]{mnras}

\usepackage{newtxtext,newtxmath}

\usepackage[T1]{fontenc}
\usepackage{ae,aecompl}

\usepackage{graphicx}	
\usepackage{amsmath}	
\usepackage{amssymb}	
\usepackage{bm}
\usepackage{multirow}
\usepackage{threeparttable}

\newcommand{\nuni}{\nu_{\mathrm{ni}}}
\newcommand{\MA}{\mathcal{M}_{\mathrm{A}}}
\newcommand{\uLA}{u_{\mathrm{LA}}}
\newcommand{\uLf}{u_{\mathrm{Lf}}}
\newcommand{\vAt}{V_{\mathrm{At}}}
\newcommand{\vAi}{V_{\mathrm{Ai}}}
\newcommand{\nuAt}{\nu_{\mathrm{At}}}
\newcommand{\nuAi}{\nu_{\mathrm{Ai}}}
\newcommand{\nuin}{\nu_{\mathrm{in}}}
\newcommand{\rhon}{\rho_{\mathrm{n}}}
\newcommand{\rhoi}{\rho_{\mathrm{i}}}
\newcommand{\xin}{\xi_{\mathrm{n}}}
\newcommand{\nH}{n_{\mathrm{H}}}
\newcommand{\mH}{m_{\mathrm{H}}}
\newcommand{\muH}{\mu_{\mathrm{H}}}
\newcommand{\gd}{\gamma_{\mathrm{d}}}
\newcommand{\nHthree}{n_{\mathrm{H},3}}
\newcommand{\uLAone}{u_{\mathrm{LA,1}}}
\newcommand{\uLfone}{u_{\mathrm{Lf,1}}}
\newcommand{\LdampA}{L_{\mathrm{damp,A}}}
\newcommand{\Ldampf}{L_{\mathrm{damp,f}}}
\newcommand{\kdampA}{k_{\mathrm{damp,A}}}
\newcommand{\kdampf}{k_{\mathrm{damp,f}}}
\newcommand{\mui}{\mu_{\mathrm{i}}}

\newcommand{\rg}{r_{\mathrm{g}}}
\newcommand{\ECR}{E_{\mathrm{CR}}}
\newcommand{\ECRz}{E_{\mathrm{CR,0}}}
\newcommand{\ECRscat}{E_{\mathrm{CR,scat}}}
\newcommand{\GCR}{\Gamma_{\mathrm{CR}}}
\newcommand{\nCR}{n_{\mathrm{CR}}}
\newcommand{\vst}{V_{\mathrm{st}}}
\newcommand{\nion}{n_{\mathrm{i}}}
\newcommand{\LA}{L_\mathrm{A}}

\newcommand{\aref}[1]{\hyperref[#1]{Appendix~\ref{#1}}}

\title[Cosmic ray transport in starbursts]{Cosmic ray transport in starburst galaxies}

\author[M. R. Krumholz et al.]{Mark R. Krumholz,$^{1,2,3,4}$\thanks{E-mail: mark.krumholz@anu.edu.au (MRK)}
Roland M. Crocker,$^{1}$
Siyao Xu,$^{5,6}$
A. Lazarian,$^{5}$
\newauthor
M. T. Rosevear$^{1}$ and
Jasper Bedwell-Wilson$^{1}$
\\
$^{1}$Research School of Astronomy and Astrophysics, Australian National University, Cotter Rd., Weston Creek, ACT 2611, Australia\\
$^{2}$ARC Centre of Excellence for All-Sky Astrophysics in Three Dimensions (ASTRO-3D), Canberra, Australia\\
$^{3}$Universit\"at Heidelberg, Zentrum f\"ur Astronomie, Institut f\"ur Theoretische Astrophysik, 69120 Heidelberg, Germany\\
$^{4}$Max Planck Institute for Astronomy, K\"onigstuhl 17, 69117 Heidelberg, Germany\\
$^{5}$Department of Astronomy, University of Wisconsin, 475 North Charter Street, Madison, WI 53706, USA\\
$^{6}$Hubble Fellow
}

\date{Accepted XXX. Received YYY; in original form ZZZ}

\pubyear{2019}

\usepackage{longtable}

\begin{document}
\label{firstpage}
\pagerange{\pageref{firstpage}--\pageref{lastpage}}
\maketitle

\begin{abstract}
Starburst galaxies are efficient $\gamma$-ray producers, because their high supernova rates generate copious cosmic ray (CR) protons, and their high gas densities act as thick targets off which these protons can produce neutral pions and thence $\gamma$-rays. In this paper we present a first-principles calculation of the mechanisms by which CRs propagate through such environments, combining astrochemical models with analysis of turbulence in weakly ionised plasma. We show that CRs cannot scatter off the strong large-scale turbulence found in starbursts, because efficient ion-neutral damping prevents such turbulence from cascading down to the scales of CR gyroradii. Instead, CRs stream along field lines at a rate determined by the competition between streaming instability and ion-neutral damping, leading to transport via a process of field line random walk. This results in an effective diffusion coefficient that is nearly energy-independent up to CR energies of $\sim 1$ TeV. We apply our computed diffusion coefficient to a simple model of CR escape and loss, and show that the resulting $\gamma$-ray spectra are in good agreement with the observed spectra of the starbursts NGC 253, M82, and Arp 220. In particular, our model reproduces these galaxies' relatively hard GeV $\gamma$-ray spectra and softer TeV spectra without the need for any fine-tuning of advective escape times or the shape of the CR injection spectrum.
\end{abstract}




\section{Introduction}

The last decade has seen a dramatic increase in our knowledge of $\gamma$-ray emission from star-forming galaxies beyond our own. At GeV energies, the \textit{Fermi}/LAT experiment has detected not only the Milky Way and its three closest star-forming companions the SMC, LMC, and M31 \citep{Abdo10c, Abdo10d, Abdo10e}, but also the more distant starburst galaxies NGC 253, NGC 4945, NGC 1068, M82, and Arp 220 \citep{Abdo10a, Nolan12a, Peng16a, Griffin16a}. Combined with ground-based data at TeV energies from HESS and VERITAS \citep{VERITAS09a, Acero09a, Abramowski12a, HESS18a}, these observations have allowed us to determine both the overall $\gamma$-ray luminosity and the spectral shape of the high energy emission from a modest sample of starburst galaxies.

The main source of $\gamma$-rays from star-forming galaxies is inelastic scattering between cosmic ray (CR) hadrons (dominantly protons) and nuclei (again, dominantly protons) in the interstellar medium (ISM), yielding neutral mesons (mostly pions) that subsequently decay into $\gamma$-rays. The $\gamma$-ray data thus far make it possible to draw a few high-level conclusions about the nature of CR proton population in starburst galaxies. One is that starburst galaxies, unlike the Milky Way or other spiral galaxies, are close to the limit of being calorimetric for protons, i.e., the majority of the energy injected into CR protons by shock acceleration is eventually converted to pions, the neutral fraction of which thence decay to $\gamma$-rays. Lines of evidence for this conclusion include both direct comparisons between $\gamma$-ray luminosities and star formation (and thus supernova) rates \citep[e.g.,][]{Lacki11a} and detailed modeling of the $\gamma$-ray spectrum \citep[e.g.,][]{Paglione12a, Yoast-Hull13a, Yoast-Hull14a, Wang18a, Peretti19a}. A second conclusion is that, again unlike in the Milky Way, CR escape from starbursts cannot be strongly energy-dependent, at least for CRs in the GeV to TeV energy range. The primary observational evidence for this conclusion is the spectral shape. The 
spectrum of the diffuse, planar $\gamma$-ray emission from the 
Milky Way is, 
in general, softer\footnote{The soft spectrum of the Galaxy's diffuse emission is, of course, a broad match to that expected given the measured distribution of the local CR population (which is also steeper than that thought to be directly injected into the Galaxy's ISM by diffusive shock acceleration).} \citep[e.g.,][]{Strong10a}
than the spectra of individual supernova remnants (SNRs) that are the main site of CR acceleration; together these observations indicate faster escape of higher energy CRs from the Galaxy. By contrast, starbursts have systematically harder $\gamma$-ray spectra than the general Milky Way, with slopes closer to those seen in individual SNRs. 

The increase in available $\gamma$-ray data has led to concomitant efforts at modelling, with the goal of inferring the primary transport and loss processes for CRs in starbursts. However, the majority of these efforts adopt a purely phenomenological model for CR transport, and attempt to infer quantities such as the diffusive or advective escape time from the spectrum. Similarly, a number of authors have simulated CRs and their role in driving winds from starburst galaxies using a variety of physical models for CR transport \citep[e.g.,][]{Booth13a, Hanasz13a, Salem13a, Girichidis16a, Ruszkowski17a}. However, in all of these simulations, transport rates and coefficients are left as free parameters that are either set based on empirical estimates from the Milky Way, or are varied in order to perform a parameter study. There has been comparatively little effort, either using analytic models or in simulations, attempting to infer the transport properties from first principles. To the extent that authors have developed either analytic \citep[e.g.,][]{Yan08a, LY14, XL18t} or numerical \citep[e.g.,][]{Kota00a, XY13} descriptions for CR transport in the highly magnetised, turbulent ISM likely to be found in starbursts, there has been been relatively little effort to use these models to study their $\gamma$-ray emission.

This is problematic because it is likely a poor assumption that the ISM processes determining CR transport in
starbursts are the same as those operating in galaxies like the Milky Way. The mid-planes of starbursts are dominated by cold, neutral gas, and the last few years have also seen a major revision in our understanding of CR transport in such media, with the primary aim of understanding how CRs propagate in the molecular clouds in our own Galaxy \citep[e.g.,][]{Xuc16, Nava16a, XLr17, XL18t}. In such clouds, the ionisation fraction is very low and, as a result, the magnetohydrodynamic (MHD) turbulence responsible for scattering CRs has a very different structure than that which prevails in ionised gas. In particular, damping of turbulence by ion-neutral drag is a first-order effect that cannot be ignored. This fundamentally changes the nature of CR transport. Our goal in this paper is to work through the implications of this revised picture for the nature of CR transport and $\gamma$-ray production in starburst galaxies.

The structure of the remainder of this paper is as follows. In \autoref{sec:cr_transport}, we examine the basic physics of CR transport in starburst galaxies by first estimating the chemical and ionisation state of the gas, and then developing a model for CR transport through gas with these properties. In \autoref{sec:gamma_rays} we use the results derived in \autoref{sec:cr_transport} to derive the implications for $\gamma$-ray production in starbursts, with particular attention to the questions of calorimetry and the spectral shape. We discuss the implications of our findings in \autoref{sec:discussion}, and summarise in \autoref{sec:conclusions}.

\section{Cosmic ray transport processes}
\label{sec:cr_transport}

Our overall goal in this section is to develop a theory for the transport of cosmic rays through the ISM of a starburst galaxy. To this end, we first estimate the bulk properties of the starburst ISM in \autoref{ssec:starburst_ISM}, and we then calculate the structure of an MHD turbulent cascade through such a medium in \autoref{ssec:cascade}. We consider the importance of the streaming instability in \autoref{ssec:streaming}, and use this to derive macroscopic diffusion coefficients in \autoref{ssec:diffusion}.

\subsection{Interstellar media in starburst galaxies}
\label{ssec:starburst_ISM}

As a first step in investigating the nature of CR transport in starburst galaxies, we investigate the structure of starbursts'  interstellar media. Starbursts are distinguished from the Milky Way primarily by their vastly higher surface and volume densities of gas, and correspondingly higher rates of star formation. We summarise some key bulk properties in the nearby starbursts Arp 220, M82, and NGC 253 in \autoref{tab:starburst_properties}. For all three galaxies, we use gas surface density measurements taken from \citet{Kennicutt98a}, adjusting the CO to mass conversion factor in these starburst systems following the recommendation of \citet{Daddi10a}. To estimate the volume density we require knowledge of the scale height or the velocity dispersion, from which we can derive the scale height. For Arp 220, direct measurements of the velocity dispersion are available from \citet{Scoville17a} and \citet{Wilson19a}. For M82 and NGC 253, no measurements of the velocity dispersion of the neutral ISM are available in the literature, and we therefore estimate the velocity dispersion, scale height, and volume density from the gas surface density $\Sigma$ and orbital period $t_{\rm orb}$ by adopting
\begin{eqnarray}
\label{eq:Qgas}
Q_{\rm gas} & \approx & \frac{2\sigma}{G \Sigma t_{\rm orb}} \\
h & \approx & \frac{\sigma^2}{\pi G \Sigma},
\end{eqnarray}
with $Q_{\rm gas} \approx 2$, following \citet{Forbes12a} and \citet{Krumholz18a}. As the table shows, even modest starbursts like NGC 253 have gas densities two orders of magnitude above the Milky Way mean ($\nH \sim 1$ cm$^{-3}$, where $\nH$ is the volume density of H nuclei; \citealt{Wolfire:2003}), and comparable to values observed in Milky Way molecular clouds; more powerful ultraluminous infrared galaxies such as Arp 220 reach mean densities four orders of magnitude above the Milky Way mean, and comparable to those found in cluster-forming clumps in the Milky Way.

\begin{table*}
    \caption{Bulk ISM properties in selected starburst galaxies}
    \label{tab:starburst_properties}
    \centering
    \begin{tabular}{llcccl}
    \hline\hline
    Parameter & Unit & \multicolumn{3}{c}{Galaxy} & References \\
    & & Arp 220 & M82 & NGC 253 & \\ \hline
    \multicolumn{6}{c}{Bulk properties}
    \\
    \hline
    Log star formation rate / area, $\log\dot{\Sigma}_*$$^{(a)}$ & $M_\odot$ pc$^{-2}$ Myr$^{-1}$ & 2.69 & 1.24 & 1.00 & K98 \\
    Log gas / area, $\log \Sigma$$^{(a)}$ & $M_\odot$ pc$^{-2}$ & $4.0$ & $2.8$ & $2.6$ &  K98 \\
    Orbital period, $t_{\rm orb}$ & Myr & 6 & 9 & 15 & K98 \\
    Velocity dispersion, $\sigma$$^{(b)}$ & km s$^{-1}$ & 100 & 25 & 26 & S17 \\
    Gas scale height, $h$$^{(b)}$ & pc & 75 & 73 & 130 & W19 \\
    Log number density, $\log n_{\rm H}$$^{(c)}$ & cm$^{-3}$ & 3.6 & 2.4 & 2.0 & \\
    $pp$ loss time, $t_\text{loss}$ & Myr & 0.011 &  0.18 & 0.44 & \\
    SNR filling factor, $f_{\rm SNR}$ &
    -- & $1.2\times 10^{-3}$ & $0.080$ & $0.095$ \\
    SNR volumetric injection rate, $\dot{\rho}_{\rm SNR}$ &
    SNe  Myr$^{-1}$  pc$^{-3}$ & 160 & 5.7 & 1.9 \\
    \hline
    \multicolumn{6}{c}{CR transport properties (derived assuming $\MA\approx 2$, see text)}
    \\ \hline
    Magnetic field, $B$ & $\mu$G & 1200 & 76 & 50 \\ [0.5ex]
    Alfv\'en damping length, $\LdampA$ & pc & 
    $0.0028 \chi_{-4}^{-3/2}$ &
    $0.023 \chi_{-4}^{-3/2}$ & 
    $0.072 \chi_{-4}^{-3/2}$ \\ [0.5ex]
    Fast damping length, $\Ldampf$ & pc & 
    $0.74 \chi_{-4}^{-3/2}$ &
    $1.8 \chi_{-4}^{-3/2}$ & 
    $4.2 \chi_{-4}^{-3/2}$ \\ [0.5ex]
    CR scattering energy, $\ECRscat^{(d)}$ & TeV &
    $730\, \chi_{-4}^{-3/2}$ &
    $370\, \chi_{-4}^{-3/2}$ &
    $760\, \chi_{-4}^{-3/2}$ &
    \\ [0.5ex]
    Streaming/Alfv\'en speed difference, $\frac{\vst}{\vAi}-1$ & 
    $10^{-3}$ &
    $5.1 \ECRz^{1.6} \chi_{-4} C_3^{-1}$ &
    $0.32 \ECRz^{1.6} \chi_{-4} C_3^{-1}$ &
    $0.078 \ECRz^{1.6} \chi_{-4} C_3^{-1}$
    \\ [0.5ex]
    Ion Alfv\'en speed, $\vAi$ & $10^3$ km s$^{-1}$ &
    $3.5 \chi_{-4}^{-1/2}$ & 
    $0.88 \chi_{-4}^{-1/2}$ & 
    $0.92 \chi_{-4}^{-1/2}$ \\ [0.5ex]
    Diffusion coefficient, $D$ & $10^{27}$ cm$^2$ s$^{-1}$ & $10.0\chi_{-4}^{-1/2}$ & 
    $2.5\chi_{-4}^{-1/2}$ &
    $4.6\chi_{-4}^{-1/2}$ \\ [0.5ex]
    CR overlap fraction, $Q_\text{CR}$ & -- & $27\chi_{-4}^{-3/4}$ & 
    $110\chi_{-4}^{-3/4}$ &
    $1200\chi_{-4}^{-3/4}$ \\
    \hline
    \multicolumn{6}{c}{CR calorimetry (derived assuming $\MA\approx 2$, see text)}
    \\ \hline
    Optical depth to CRs, $\tau_{\rm eff}$ & -- & $7.2\chi_{-4}^{1/2}$ & $1.8\chi_{-4}^{1/2}$ & $1.1\chi_{-4}^{1/2}$ \\
    CR energy density normalisation, $\Phi_0$$^{(e)}$ & $10^{-8}$ erg cm$^{-3}$ & $1.8\chi_{-4}^{1/2}$ & $0.26\chi_{-4}^{1/2}$ & $0.14\chi_{-4}^{1/2}$ \\
    Ratio of streaming to diffusive speed, $\Upsilon_0$ & -- & 10.7 & 10.7 & 10.7 \\
    Midplane CR energy density, $U(0)$$^{(e)}$ & keV cm$^{-3}$ &
    4.6 & 1.4 & 1.1 \\
    Calorimetric fraction, $f_{\rm cal}$ & -- & 0.99 & 0.91 & 0.83 \\
    \hline\hline
    \end{tabular}
    \begin{tablenotes}
    \item In this table, $\chi_{-4}$ is the ion mass fraction normalised by $10^{-4}$, $\ECRz$ is the CR energy in units of GeV, and $C_3$ is the CR energy density normalised to 1000 times the Solar neighbourhood value. References are as follows: K98 = \citet{Kennicutt98a},
    S17 = \citet{Scoville17a}, W19 = \citet{Wilson19a}
    \item $^{(a)}$ Gas surface densities taken from \citet{Kennicutt98a} have been adjusted to $\alpha_{\rm CO}=0.8$ $M_\odot$ pc$^{-2}$ / K km s$^{-1}$, and star formation surface densities have been homogenised to a \citet{Chabrier05a} IMF, as recommended by \citet{Daddi10a}
    \item $^{(b)}$ Only observed for Arp 220; for M82 and NCG 253, derived as discussed in the main text
    \item $^{(c)}$ Derived from the scale height and gas surface density: $n_{\rm H} = \Sigma/\muH m_{\rm H} h$, where $\muH=1.4$ is the mean mass per H nucleon and $m_{\rm H}=1.67\times 10^{-24}$ is the hydrogen mass
    \item $^{(d)}$ Computed for pitch angle $\alpha=45^\circ$; results scale as $\propto 1/\sin\alpha$ 
    \item $^{(e)}$ These quantities are upper limits, derived in the case of zero advective escape.
    \end{tablenotes}
\end{table*}

A second and related characteristic of the starbursts, as discussed by \citet{Murray10a} and \citet{Lacki13a}, is that the molecular material is likely volume-filling near the midplane. In the Milky Way midplane, hot and warm ionised gas occupy $\sim 50\%$ of the available volume \citep[e.g.,][]{Drai11}. In contrast, in a starburst there is no diffuse warm ionised medium due to the very short mean free path of ionising photons. The volume filling factor of supernova remnants (SNRs) is $f_{\rm SNR}\approx 1-e^{-Q_{\rm SNR}}$ \citep{McKee77a, Heckman90a, Lacki13a}, where \begin{eqnarray}
    \lefteqn{Q_{\rm SNR} \approx 3.4 E_{51}^{1.28}}
    \nonumber \\
    & &  \left(\frac{\rho_{\rm SN}}{10\;\mathrm{kpc}^{-3}\;\mathrm{yr}^{-1}}\right) \left(\frac{\nH}{100\;\mathrm{cm}^{-3}}\right)^{-0.14} \left(\frac{P/k_{\rm B}}{10^7\;\mathrm{K}\;\mathrm{cm}^{-3}}\right)^{-1.3}.
\end{eqnarray}
Here $E_{\rm 51}$ is the mechanical energy released per core collapse supernova, $\rho_{\rm SN}$ is the density of supernovae, and $P$ is the gas pressure. The density of supernovae near the midplane is $\rho_{\rm SN} \approx \dot{\Sigma}_*/h m_{\rm SN}$, where $\dot{\Sigma}_*$ is the star formation rate per unit area, and $m_{\rm SN}\approx 100$ $M_\odot$ is the mean mass of stars that must be formed to eventually yield one supernova. The pressure in turn must be related to the gas surface density by
\begin{equation}
    P = \pi G f_{\rm gas}^{-1} \Sigma^2,
\end{equation}
where $f_{\rm gas}$ is the gas fraction at the midplane. Evaluating $f_{\rm SNR}$ using the parameters given in \autoref{tab:starburst_properties} and assuming a gas fraction $f_{\rm gas}=0.5$ gives $f_{\rm SNR}\approx 10^{-3}$ in Arp 220, and $f_{\rm SNR} < 0.1$ in M82 and NGC 253, suggesting that the midplanes of these galaxies cannot be filled with supernova-heated gas.\footnote{Our conclusion regarding Arp 220 is the same as that of \citet{Lacki13a}, but our conclusions about M82 and NGC 253 differ. Our formulae are identical to \citeauthor{Lacki13a}'s, so the sole source of difference lies in the input data -- \citeauthor{Lacki13a} adopts fixed gas depletion times of 20 Myr and scale heights of 50 pc for these two galaxies, whereas the measured values listed in \autoref{tab:starburst_properties} give roughly half the star formation rate and $1.5-2$ times the scale height. \citet{Lacki13a} also argues that the relatively fast expansion speeds measured for SNRs in M82 and NGC 253 (as opposed to those in Arp 220) favour volume-filling hot phase. However, this conclusion is based on the analytic expansion models of \citet{Chevalier01a}, which assume as uniform density ambient medium. More recent numerical work by \citet{Martizzi15a} shows that expansion velocities are larger in a turbulent medium; using their numerical expansion rates rather than \citeauthor{Chevalier01a}'s analytic ones removes any conflict between a model with a volume-filling neutral medium and the measured expansion speeds in NGC 253 and M82. That said, the sensitivity of the result to these factor of two level differences does suggest that one should treat the result with some caution. In particular, our estimate of $Q_{\rm SNR}$ does not include the effects of faster expansion in a turbulent medium, which would tend to raise $Q_{\rm SNR}$. Conversely, it does not include the effects of SN clustering, which would tend to lower it.} Note that this does not mean that SNRs cannot break out of the galaxy and drive a wind \citep[e.g.,][]{Strickland09a}, simply that the hot gas must escape through chimneys that occupy much less than order unity of the available volume within one scale height of the midplane.

The presence of a volume-filling neutral gas at the midplane of starbursts has important implications for CR propagation. To investigate the properties of the medium through which CRs propagate in starbursts, we make use of the chemistry and radiative transfer code \textsc{despotic} \citep{Krumholz14b}. For each of the starburst galaxies listed in \autoref{tab:starburst_properties}, we use \textsc{despotic} to calculate the equilibrium thermal and chemical state of a medium with volume density, surface density, and gas velocity dispersion equal to the value given in the Table. We perform the chemical calculation using the reduced H-C-O chemical network of \citet{Gong17a}. Our thermal calculation includes heating by CR ionisation and the photoelectric effect, cooling by rotational and fine structure line emission from CO, $^{13}$CO, C, C$^+$, and O excited by collisions with H, H$_2$, H$^+$, He, and $e^-$, and dust-gas energy exchange. We compute level populations using an escape probability method, with escape probabilities computed using \textsc{despotic}'s spherical geometry option. Full details on the assumptions and methods of the calculation are given in \citet{Krumholz14b}. \textsc{Despotic} makes use of data from the Leiden Atomic and Molecular Database (LAMDA; \citealt{Schoier05a}).\footnote{Data for collision rate coefficients used in the calculation come from the following sources: CO and $^{13}$CO: \citet{Yang10a}; C: \citet{Launay77a, Johnson87a, Roueff90a, Schroder91a, Staemmler91a}; C$^+$: \citet{Launay77a, Flower77a, Flower88a, Barinovs05a, Lique13a, Wiesenfeld14a}; O: \citet{Launay77a, Chambaud80a, Jaquet92a, Bell98a, Abrahamsson07a}.
}

For the purposes of this calculation we must adopt values for four parameters, though three have only minor effects. The most significant parameter is the rate of primary ionisations due to CRs, which is very poorly constrained in starburst galaxies \citep[e.g.,][]{Papadopoulos10a, Bisbas15a, Narayanan17a}. A plausible upper limit can be obtained by noting that the most luminous of our sample starbursts, Arp 220, has a $\gamma$-ray luminosity $\approx 10^3$ times that of the Milky Way \citep{Griffin16a}, combined with a somewhat smaller gas mass \citep{Scoville97a}. Thus to the extent that the low-energy CRs that produce ionisation behave similarly to their higher-energy cousins responsible for $\gamma$-ray emission, we expect Arp 220's CR ionisation rate to be $\sim 10^3-10^4$ times that of the Milky Way. The other starbursts are intermediate. Given this discussion, we consider a wide range of primary ionisation rates, ranging from typical Milky Way values ($\zeta \approx 10^{-16}$ s$^{-1}$ -- \citealt{Indriolo12a}) up to values $10^4$ times larger.

The remaining three parameters are the dust temperature, the interstellar radiation field (ISRF) intensity, and the relative abundances of C, $^{13}$C, and O relative to H. Dust temperature affects the rate of dust-gas energy exchange. This is sub-dominant compared to CR heating and line cooling at the densities with which we are concerned, but can change the gas temperature by $\sim 1$ K. Dust temperatures near starburst midplanes are poorly constrained, but are almost certainly higher than found in the Milky Way. We adopt $T_d = 80$ K for Arp 220, and $T_d = 40$ K for M82 and NGC 253. The ISRF affects the thermal state via the photoelectric effect, and the chemical state via photodissocation reactions. As with the dust temperature, the effects of this choice are small because the high optical depths implied by the large surface densities of starburst galaxies means that the ISRF experienced by most of the gas is greatly reduced compared to the unshielded value. Following \citet{Krumholz13c}, we estimate the unshielded ISRF strength by scaling the Milky Way ISRF by the ratio of star formation surface densities between the starbursts and the Milky Way, using the values given in \autoref{tab:starburst_properties}. Finally, the chemical abundances are known to be non-Solar in starburst regions. Since direct measurements in starbursts are few, we take these abundances to be equal to the Galactic Centre values given by \citet{Wilson92a}.

\begin{figure}
    \centering
    \includegraphics[width=\columnwidth]{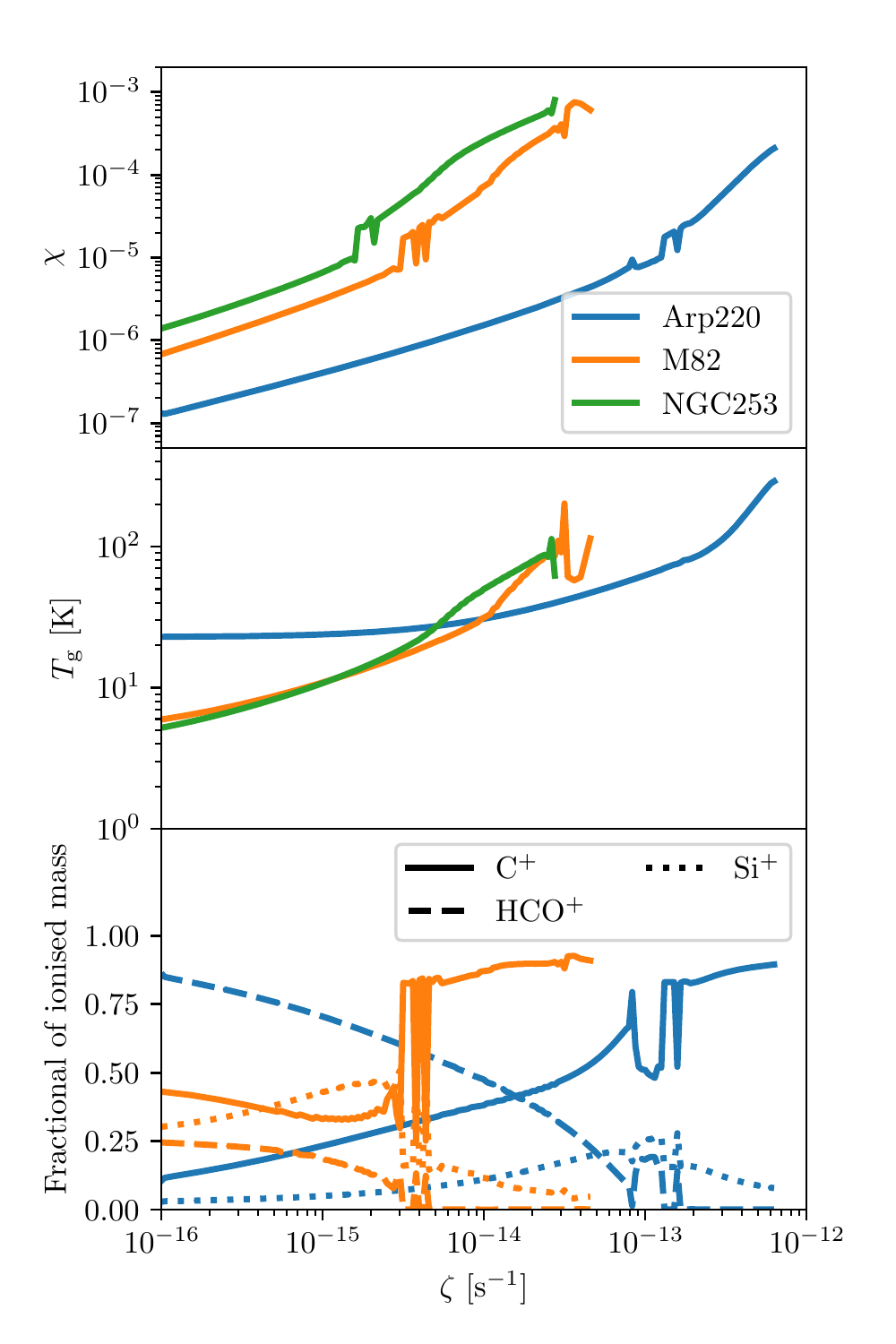}
    \caption{Results of a calculation of the chemical and thermal equilibria for the ISM in starburst galaxies as a function of primary cosmic ray ionisation rate $\zeta$ using \textsc{despotic} \citep{Krumholz14b}. The three colours correspond to calculations using our fiducial gas volume density, column density, and velocity dispersions for Arp 220, NGC 253, and M82 (colours, as indicated in the legend) given in \autoref{tab:starburst_properties}. The top panel shows the ionised gas mass fraction $\chi$, the middle shows the equilibrium gas temperature $T_{\rm g}$, and the bottom shows the fraction of the ionised gas in the three dominant ionised species, C$^+$, HCO$^+$, and Si$^+$; in the bottom panel we omit NGC 253 for clarity, since it is substantially similar to M82. We truncate the lines at the point where, for larger values of $\zeta$, the equilibrium ISM state is warm and atomic ($T_{\rm g} > 300$ K) rather than cold and molecular, inconsistent with observations, which show that in starbursts the gas mass is dominated by cool molecular material. Note that the rapid and non-monotonic variation in the equilibrium chemical state apparent at some values of $\zeta$ is a real effect, which occurs because, for complex chemical networks, transitions between different chemical regimes (e.g., C$^+$-dominated versus HCO$^+$-dominated) are chaotic.}
    \label{fig:ionstate}
\end{figure}

We show the resulting equilibrium ionised mass fractions $\chi$, gas temperatures, and the fraction of the ionised mass in various species in \autoref{fig:ionstate}. We find that, for plausible CR ionisation rates, the gas in starbursts will be relatively warm compared to a Milky Way molecular clouds (temperature tens of K), and will have an ionised mass fraction of order $\chi \sim 10^{-4}$. Unlike in Milky Way molecular clouds, where HCO$^+$ provides most of the ionised mass, in starbursts at higher ionisation rates the predominant ion by mass shifts to  C$^+$, which is produced by the reaction chain
\begin{eqnarray}
\mathrm{He} + \mathrm{CR} & \rightarrow & \mathrm{He}^+ + e^- + \mathrm{CR} \\
\mathrm{He}^+ + \mathrm{CO} & \rightarrow & \mathrm{He} + \mathrm{O} + \mathrm{C}^+.
\end{eqnarray}
HCO$^+$ and Si$^+$ also make a subdominant contribution to the ions.

\subsection{Turbulent cascades and cosmic ray scattering}
\label{ssec:cascade}

In the Milky Way, the dominant CR transport process is (or, to a reasonable approximation, can be treated as though it were) diffusion caused by resonant interactions between CRs and turbulent fluctuations of magnetic fields mainly on size scales comparable to the CR gyroradius \citep{XL18t}\footnote{These fluctuations may, depending on the ISM conditions and the CR energy, be dominantly self-excited, i.e., generated by the CRs themselves via the streaming instability, or be due to extrinsic turbulence cascading down to gyroradius scales.}. We now show that this mechanism must break down in starburst-like environments due to their low ionisation fractions, and instead a different transport regime prevails. We begin this section with a qualitative summary of turbulent cascades in partially-ionised media, following the recent review by \citet{XLr17}; we refer readers there for more detail.

We first adopt the wave description of magnetic fluctuations to consider the damping process. 
In weakly-ionised media, collisions of a neutral with ions occur infrequently due to the low space density of ions. This has important implications for the propagation of MHD waves. Consider a medium in which the frequency of neutral-ion collisions is $\nuni$. As a wave with frequency $\nu \ll \nuni$ propagates, many collisions will occur per oscillation, which will force the ions and neutrals to move together, acting as a single magnetised fluid supporting the usual family of MHD waves. By contrast, a wave with $\nu \gg \nuin$, where $\nuin$ is the frequency of ion-neutral collisions, will see essentially no ion-neutral collisions during each oscillation period, and as a result the medium acts like two completely separate and decoupled fluids. MHD waves will propagate only in the ions, while separate and decoupled sound waves will propagate in neutrals when the frequency of sound waves exceeds $\nuni$. Most importantly for our purposes,  Alfv\'{e}n waves at intermediate frequencies in the range $\nuni < \nu <  \nuin$ are cut off
\citep{Kulsrud_Pearce}: ions attempt to oscillate in response to perturbations in the magnetic field, but still collide with the surrounding neutrals. 
Neutrals, which are essentially decoupled from ions due to their infrequent collisions with ions, cannot move  
with the Alfv\'{e}n waves. 
The ion-neutral collisions 
prevent the ions from oscillating freely. Thus the ion-neutral collisions will convert organised Alfv\'{e}n wave motions in the weakly coupled ions and neutrals into microscopic random motions, dissipating them into heat much like viscosity does in non-magnetic fluids. No comparable damping occurs for the sound waves propagating in the decoupled neutrals. 
In the case when the above ion-neutral collisional damping dominates over the neutral viscous damping, 
the sound waves in neutrals are damped at the viscous scale due to the viscosity in neutrals, which is smaller than the ion-neutral collisional damping scale \citep{Xuc16}. 

MHD turbulence has an energy cascade from large to small length scales. 
In the case of strong Alfv\'{e}nic turbulence \citep{GS95, LV99}, 
turbulent motions perpendicular to the local magnetic field 
and wave-like motions along the magnetic field 
are dynamically coupled with the same timescales. 
The above-mentioned damping of Alfv\'{e}n waves at intermediate frequencies acts as a barrier to the Alfv\'{e}nic turbulent cascade: large-scale, low-frequency motions occur in a coupled ion-neutral fluid and cascade down to smaller scales. In the decoupled neutrals the hydrodynamic cascade goes down to the viscous scale \citep{XLY14,BurL15}. 
However, in the weakly coupled ions and neutrals and the magnetic field, the cascade stops at the length scale corresponding to the wave frequencies that are damped by ion-neutral collisions. 
Note that the damping scale of the Alfv\'{e}nic turbulent cascade and that of the Alfv\'{e}n waves are the 
perpendicular and parallel length scales with respect to the local magnetic field. 
Their scaling relation is determined by the scale-dependent anisotropy of MHD turbulence \citep{CL02_PRL}. 
Below these scales, large-scale motions of the fluid induce no corresponding perturbations in the ions or the magnetic field.

We now proceed to evaluate this damping scale quantitatively, following the treatment given in \citet{XLY14,Xuc16} and \citet{XLr17}. Compressible MHD turbulence can be decomposed into Alfv\'{e}nic, slow, and fast modes \citep{CL02_PRL,CL03}. We first consider Alfv\'enic modes; since they carry most of the energy of the MHD cascade and, in fact, slow modes are passively mixed by Alfv\'{e}nic turbulence and follow the same cascade as Alfv\'{e}nic modes
\citep{LG01,CL03}. For the latter, the damping scale depends on whether the Alfv\'en Mach number of the turbulence at its injection scale, $\MA = \uLA/\vAt$, is larger or smaller than unity; here $\uLA$ is the turbulent velocity of Alfv\'{e}nic modes at the injection scale $L$ of the turbulence, 
$\vAt = B /\sqrt{4\pi \rho}$ is the Alfv\'{e}n speed in terms of the total mass density $\rho$ ($=\nH\muH\mH$), $\muH$ is the mean mass per H nucleus in units of hydrogen masses $\mH$ ($\approx 1.4$ for standard cosmic composition), and $B$ is the strength of mean magnetic field. Unfortunately computing $\MA$ is not easy. If turbulent energy were in equipartition between Alfv\'enic, fast, and slow modes, we would have $\uLA\approx \sigma/\sqrt{3}$, where $\sigma$ is the ISM velocity dispersion in the galaxy, but since the Alfv\'enic modes are non-compressive and the other two are compressive, there is likely somewhat more energy in the Alfv\'enic modes; we shall adopt $\uLA\approx\sigma/\sqrt{2}$ as a fiducial choice, corresponding to half the energy being in Alfv\'enic modes. However, magnetic field strengths, and thus Alfv\'en speeds, are only poorly known in the bulk ISM of starburst galaxies. Direct measurements are only available in the ionised outflows \citep[e.g.,][]{Heesen11a, Adebahr13a} or in regions of maser emission that are likely unrepresentative of mean conditions in the ISM \citep[e.g.,][]{McBride15a}.

In the absence of strong observational constraints, we turn to theory. As argued by \citet[see also \citealt{Thompson06a}]{Lacki13a}, the magnetic fields in starbursts are likely the result of a turbulent dynamo that converts some fraction of the kinetic energy of turbulence to magnetic energy. There have been extensive studies of the turbulent dynamo in solenoidal turbulence at different ionization fractions \citep{XL16,XLr17,Xub19} and in compressible turbulence \citep[e.g.][]{Fede11,Federrath14b,Federrath16a} in recent years. For high Reynolds number and magnetic Prandtl number, as expected for astrophysical plasmas, and for turbulent driving that is predominantly solenoidal, as expected for turbulence in starbursts that is driven mainly by galactic shear and gravity \citep{Krumholz18a}, the simulations of \citet{Federrath16a} suggest that the dynamo saturates at a ratio of magnetic to kinetic energy density $E_{\rm B}/E_{\rm K}\sim 0.05-0.1$. For our fiducial assumption that half the energy is in Alfv\'en modes, this corresponds to $\MA\approx 2$, and we shall adopt this as our fiducial choice. However, we caution that the exact value may be sensitive to the details of how the turbulence is driven, since \citet{XL16} shows that the cascade that drives the dynamo starts at scales that are slightly below the turbulence driving scale, and thus the exact ratio of $E_{\rm B}$ to $E_{\rm K}$ depend on the amount of energy stored in the very largest modes. We will therefore also consider variations about this value. We list the magnetic field strengths that correspond to our fiducial choice in \autoref{tab:starburst_properties}, and note that they are within a factor of a few of those inferred for ionised or masing gas in the corresponding galaxies \citep{Heesen11a, Adebahr13a, McBride15a}, and thus are not \textit{a priori} unreasonable.

The characteristic wavenumber at which Alfv\'en modes damp is given by \citep{XLr17}
\begin{equation}
    \kdampA = \left(\frac{2\nuni}{\xin}\right)^{3/2} L^{1/2} \uLA^{-3/2} \max\left(1, \MA^{-1/2}\right),
\end{equation}
where $\xin = \rhon / \rho = 1-\chi \approx 1$ is the neutral mass fraction and $\rhon$ is the neutral mass density. Compared with Alfv\'{e}nic and slow modes, fast modes have a slower cascade and are usually more severely damped. The corresponding damping wavenumber for fast modes is
\begin{equation}
    \kdampf = \left(\frac{2\nuni}{\xin}\right)^{2/3} L^{-1/3} \uLf^{4/3} \vAt^{-2} = 
    \left(\frac{2\nuni}{\xin}\right)^{2/3} L^{-1/3} \frac{\uLf^{4/3}}{\uLA^2} \MA^2
\end{equation}
where $\uLf$ is the amplitude of the turbulent velocity in fast modes at size scale $L$. 

The damping length scales for these modes are $\LdampA = 2\pi/\kdampA$ and similarly for $\Ldampf$ and $\kdampf$. To see how this depends on ionisation fraction, we note that the neutral-ion collision frequency is $\nuni = \gd \rhoi$, where $\rhoi = \chi \rho$ is the ion density and $\gd\approx 4.9 \times 10^{13}$ cm$^3$ g$^{-1}$ s$^{-1}$ is the ion-neutral drag coefficient \citep{Drai83,Shu92}\footnote{In principle $\gd$ should depend weakly on the chemical mix of ionised species; however, since this effect is very weak, we ignore it here for simplicity.}. Thus
\begin{eqnarray}
    \LdampA & = & \frac{\pi}{\sqrt{2L}} \left(\frac{\uLA}{\gd \chi \rho}\right)^{3/2} \min\left(1, \MA^{1/2}\right) 
    \nonumber
    \\ 
    & \approx & \frac{0.0011}{L_2^{1/2}} \, \left(\frac{\uLAone}{\nHthree \chi_{-4} }\right)^{3/2} \min\left(1, \MA^{1/2}\right) \;\mathrm{pc} 
    \label{eq:LdampA}
    \\
    \Ldampf & = & \pi \frac{\uLf^2}{\uLA^{4/3}} 
    \left(\frac{\sqrt{2L}}{\gd\chi\muH\mH\nH}\right)^{2/3} \MA^{-2}
    \nonumber \\
    & \approx & 1.7 \frac{\uLfone^2}{\uLAone^{4/3}} \left(\frac{\sqrt{L_2}}{\nHthree\chi_{-4}}\right)^{1/3} \MA^{-2} \;\mathrm{pc}
\end{eqnarray}
where $L_2 = L/100$ pc, $\uLAone = \uLA/10$ km s$^{-1}$, $\nHthree = \nH/10^3$ cm$^{-3}$, $\chi_{-4} = \chi/10^{-4}$, and $\uLfone=\uLf/10$ km s$^{-1}$. 

To evaluate these length scales for our sample galaxies, we take the outer scale of the turbulence to be $L = h$ (i.e., we assume turbulence is driven on scales of the galactic scale height), and we assume that the turbulence on size scale $L$ is in equipartition between fast, slow, and Alfv\'en modes (so that $\uLA = \uLf = \sigma/\sqrt{3}$). We report our results in \autoref{tab:starburst_properties}. The primary lesson from these numerical results is that, for the characteristic ionisation fractions we have derived, Alfv\'enic and slow modes in starburst interstellar media damp at scales $\sim 10^{-3}-10^{-2}$ pc, while fast modes damp at scales of $\sim 0.1 - 1$ pc.

It is informative to compare these length scales to the gryoradii of CRs moving in the galactic magnetic field,
\begin{equation}
    \rg = \frac{\gamma m c v \sin\alpha}{e B} \approx \frac{\ECR \sin\alpha}{e B} \sim 10^{-6} \  
    \ECRz \ B_0^{-1} \ {\rm pc},
    \label{eq:rg}
\end{equation}
where $\ECRz = \ECR/1$ GeV, $B_0 = B/\mu$G, and $\gamma$, $m$, $v$, $\alpha$, and $\ECR$ are the Lorentz factor, mass, velocity, pitch angle, and kinetic energy of the CR, and in the second step we have assumed that CRs are highly-relativistic, $\gamma \gg 1$ and $v\approx c$. Since cosmic rays scatter resonantly only off magnetic field fluctuations on scales $\sim \rg$, if $\kdampA \rg \ll 1$, then the externally-driven turbulence will be unable to scatter CRs effectively. (We consider $\kdampA$ rather than $\kdampf$ because $\kdampA \gg \kdampf$.) We can use the condition $\kdampA \rg = 1$ to define a critical CR energy $\ECRscat$ below which CRs will not be effectively scattered by turbulence injected at a large scale. This is
\begin{eqnarray}
    \ECRscat & = & 
    \sqrt{\frac{\pi \uLA^5}{2 L \gd^3 \chi^3}} \frac{e}{\muH \mH \nH \sin\alpha} \min\left(\MA^{-1},\MA^{-1/2}\right)
    \nonumber \\
    & \approx & 27
    \frac{\uLAone^{5/2}}{L_2^{1/2} \chi_{-4}^{3/2} \nHthree}
    \frac{\min\left(\MA^{-1}, \MA^{-1/2}\right)}{\sin\alpha} \;\mathrm{TeV}.
\end{eqnarray}
Note that we have rewritten the magnetic field $B$ in terms of the Alfv\'en Mach number $\MA$ for convenience. We evaluate $\ECRscat$ numerically for the parameters appropriate to our galaxy sample in \autoref{tab:starburst_properties}. We find that in starburst galaxies the minimum energies for efficient scattering are of order hundreds of TeV, so scattering by externally-driven turbulence in the ISM is unimportant for nearly all CRs \citep{Xuc16,XL18t}.

\subsection{Streaming instability}
\label{ssec:streaming}

Given that CRs cannot be scattered effectively by the MHD turbulent cascade in the ISM of a starburst galaxy, we next show that they will instead interact with Alfv\'en waves that they themselves generate via the streaming instability \citep{Lerche,Kulsrud_Pearce,Wen69,Ski71}, and this will ultimately limit their propagation speed. The streaming instability occurs when the CRs perturb the magnetic field, generating Alfv\'{e}n waves at a wavelength comparable to $\rg$. Our first step is to determine the coupling regime in which these waves propagate, i.e., whether the Alfv\'en waves generated by the streaming instability propagate in the ions and magnetic field only, or if there will be significant cross-talk with the neutrals.

The coupling regime is determined by the ratio of the Alfv\'en wave frequency to the collision frequency \citep{Kulsrud_Pearce}. Alfv\'en waves of wavelength $\rg$ propagating in the combined ion-neutral fluid will have frequency $\nuAt = \vAt/\rg$, and the ratio of this to the neutral-ion collision rate is
\begin{eqnarray}
R_1 & = & \frac{\nuAt}{\nuni}
= \frac{2 e \sqrt{\pi} \uLA^2}{\gd\chi\MA^2 \ECR \sqrt{\muH\mH\nH}\sin\alpha}
\nonumber \\
& = & 
4.5\times 10^6 \, \frac{\uLAone^2}{\ECRz \nHthree^{1/2}\chi_{-4} \MA^2\sin\alpha} \, .
\end{eqnarray}
Similarly, for Alfv\'en waves propagating solely in the ions, the frequency for waves of wavelength $\rg$ is $\nuAi=\vAi/\rg$, where $\vAi = B/\sqrt{4\pi\rhoi}$ is the ion Alfv\'en speed. The ion-neutral collision rate is $\nuin = \gd \rhon$, and thus the ratio of Alfv\'en frequency to collision frequency for purely ionic waves is
\begin{eqnarray}
R_2 & = & \frac{\nuAi}{\nuin} = \chi^{1/2} R_1
\nonumber \\
& = & 4.5\times 10^4 \, \frac{\uLAone^2}{\ECRz \nHthree^{1/2}\chi_{-4}^{1/2} \MA^2\sin\alpha}.
\end{eqnarray}
We see that, for the conditions characteristic of starburst galaxies, we will have $R_1 \gg R_2 \gg 1$, indicating that the Alfv\'en waves generated by the resonant streaming instability are in the decoupled regime, where collisions occur much less frequently than once per oscillation period and thus ions and neutrals are mostly decopuled.

In the decoupled regime, the rate at which CRs are able to stream is set implicitly by the condition that, at their streaming speed, the rate at which they drive Alfv\'en waves via the streaming instability balances the rate at which those waves dissipate due to ion-neutral damping. The only other significant dissipation mechanism for the streaming instability, turbulent damping, is comparatively unimportant in starburst-like environments; we demonstrate this in \aref{app:turb_damping}, and for this reason do not discuss turbulent damping further in the main text. The condition that growth of Alfv\'en waves via streaming instability balance ion-neutral dissipation is an attractor, in the sense that, if the CRs stream at less than this speed, damping will sap the Alfv\'en waves, which in turn will reduce CR scattering and allow them to stream faster; conversely, if the CRs are travelling at above the speed that satisfies this condition, the amplitude of the Alfv\'en waves they produce will grow, scattering them more effectively and reducing their streaming speed. Thus we must balance growth against damping. The damping rate in the decoupled regime is \citep{Kulsrud_Pearce}
\begin{equation}
    \omega_{\mathrm{d}} = \frac{\nuin}{2},
    \label{eq:omegad}
\end{equation}
while the growth rate of the streaming instability is
\citep{Kulsrud_Pearce}
\begin{equation}
   \GCR = \frac{e B}{m c} \frac{\nCR (> \gamma)}{\nion} \left(\frac{\vst}{\vAi}-1\right),
   \label{eq:Gamma_CR}
\end{equation}
where $\vst$ is the CR streaming velocity, $\nion=\chi\rho/\mui \mH$ is the ion number density, $\mui$ is the mean atomic mass per ion, and 
$\nCR(>\gamma)$ is the number density of CRs with the Lorentz factor larger than $\gamma$, which is $\gamma = 1.0675$ for CR protons at $1$ GeV.

If the CRs have a powerlaw distribution of Lorentz factors $d\nCR/d\gamma\propto \gamma^{-p}$, then we can write $\nCR(>\gamma) = C \gamma^{-p+1}$; for the Milky Way near the Solar Circle, $C = C_\mathrm{MW} \approx 2\times 10^{-10}$ cm$^{-3}$ and $p\approx 2.6$ \citep{Wentzel74,FG04}. Adopting this functional form for $\nCR(>\gamma)$ and setting $\GCR=\omega_{\mathrm{d}}$, we have
\begin{eqnarray}
\frac{\vst}{\vAi}-1 & = & 
\frac{\gd \chi \MA c}{4 C e \uLA \mui \gamma^{-p+1}} \sqrt{\frac{\mH m^2 \muH^3 \nH^3}{\pi}}
\nonumber \\
& = & 2.3\times 10^{-3} \frac{\ECRz^{p-1} \nHthree^{3/2} \chi_{-4}\MA}{C_3 \uLAone},
\label{eq:vratio}
\end{eqnarray}
where $C_3 = C/1000 C_\mathrm{MW} = C/2\times 10^{-7}$ cm$^{-3}$, and the numerical evaluation is for CR protons ($m=\mH$) and a population of ions dominated by C$^+$ ($\mui=12$). Thus unless the CR energy density in starbursts is so small as to be comparable to that in the Milky Way ($C_3 \sim 10^{-3}$), the streaming velocity will be very close to $\vAi$, the ion Alfv\'en speed, i.e.,
\begin{equation}
    \vst \approx \vAi \approx 1000 \frac{\uLAone}{\chi_{-4}^{1/2} \MA} \; \mathrm{km}\;\mathrm{s}^{-1}.
    \label{eq:vstream}
\end{equation}
We give expected values of $\vst/\vAi-1$ and $\vAi$ for our example starbursts in \autoref{tab:starburst_properties}. Note that one implication of our result is that the CR streaming velocity is nearly independent of energy up to CR energies of $\sim 1$ TeV. It is worth noting that \citet{Yan08a} obtained a similar result -- that the CR transport rate is nearly energy-independent -- for fully ionised gas as a result of damping of fast modes.

\subsection{Macroscopic diffusion coefficients}
\label{ssec:diffusion}

Having determined that the dominant mode of CR transport is via streaming along field lines as $\vst\approx \vAi$, our final step is to use this information to estimate the effective macroscopic diffusion coefficient for CR transport. Our discussion here follows that in \citet{Yan08a}. Before beginning our calculation, however, we should clarify the physical picture that motivates us to describe this process in terms of a ``macroscopic'' diffusion coefficient. If the turbulence in a starburst disc were highly sub-Alfv\'enic, so that the field lines were rigid, and those field lines were oriented largely out of the disc, then it would be completely straightforward to calculate how long CRs take to escape: they would stream straight along the field lines at $\vst$, and thus the time to escape would just be $\sim h/\vst$, where $h$ is the disc scale height. However, the turbulence in a starburst disc is not in fact strongly sub-Alfv\'enic. Consequently, the field lines are tangled rather than straight, and are also constantly being re-arranged by the turbulence. Thus even though CRs stream along field lines, when viewed on scales larger than the coherence length of the field lines, the process of CR transport nonetheless resembles diffusion due to field line random walk (FLRW).

We describe FLRW in terms of a ``macroscopic'' diffusion coefficient to indicate two things. First, we mean that we are averaging over scales larger than the coherence length of the MHD turbulence, so that field lines are tangled and moving rather than fixed and rigid on the scales with which we are concerned. Second, we mean macroscopic to distinguish FLRW from the process of transport via diffusion along field lines. The transport equation for CRs includes both a term describing streaming of CRs along with their confining Alfv\'en waves, which arises from the first-order term in distribution of CR pitch angles, and a term describing diffusion of CRs relative to their bulk flow, which arises from the second-order term \citep[e.g.,][]{Zweibel17a}; we refer to the process described by the second-order term as microscopic diffusion, since it occurs on scales of the CR gyroradius, rather than on scales of the coherence length of the magnetic field. The microscopic diffusion coefficient is energy-dependent, but we focus on macroscopic rather than microscopic diffusion because both analytic calculations \citep{Ski71} and recent numerical results \citep{Wiener17a} show that streaming is a significantly more important transport mechanism than microscopic diffusion. Consequently, we neglect microscopic diffusion and its energy-dependence in comparison to streaming for the purposes of computing the macroscopic diffusion coefficient.

Having understood the goal of our calculation, we now proceed to make it. The diffusion rate due to FLRW is determined by the coherence length of the field, which is related to the injection length of the turbulence by \citep{Brunetti_Laz}
\begin{equation}
    \LA \approx L \min\left(1, \MA^{-3}\right).
\end{equation}
The corresponding diffusion coefficient for CRs in the direction parallel to the large-scale field is
\begin{eqnarray}
D_{\parallel} & \approx & \vst \LA
 \\
& \approx & 3.1\times 10^{28} \frac{\uLAone L_2}{\sqrt{\chi_{-4}}} \min\left(\MA^{-1},\MA^{-4}\right) \;\mathrm{cm}^2\;\mathrm{s}^{-1}, \nonumber
\label{eq:dpar}
\end{eqnarray}
where in the second line we have evaluated $\vst$ in the low-energy limit where $\vst\approx \vAi$. If $\MA > 1$, as we assume in our fiducial case, then the perturbations in the field are not preferentially aligned with the large-scale mean field, and thus the  diffusion coefficient perpendicular to the large-scale field is the same as that parallel to it, $D_\perp \approx D_\parallel$, so that there is a single diffusion coefficient $D$ in all directions. For $\MA < 1$, the perpendicular diffusion coefficient is smaller than the parallel one by a factor of $\MA^4$ \citep{Yan08a, XY13}.  

\begin{figure}
    \centering
    \includegraphics[width=\columnwidth]{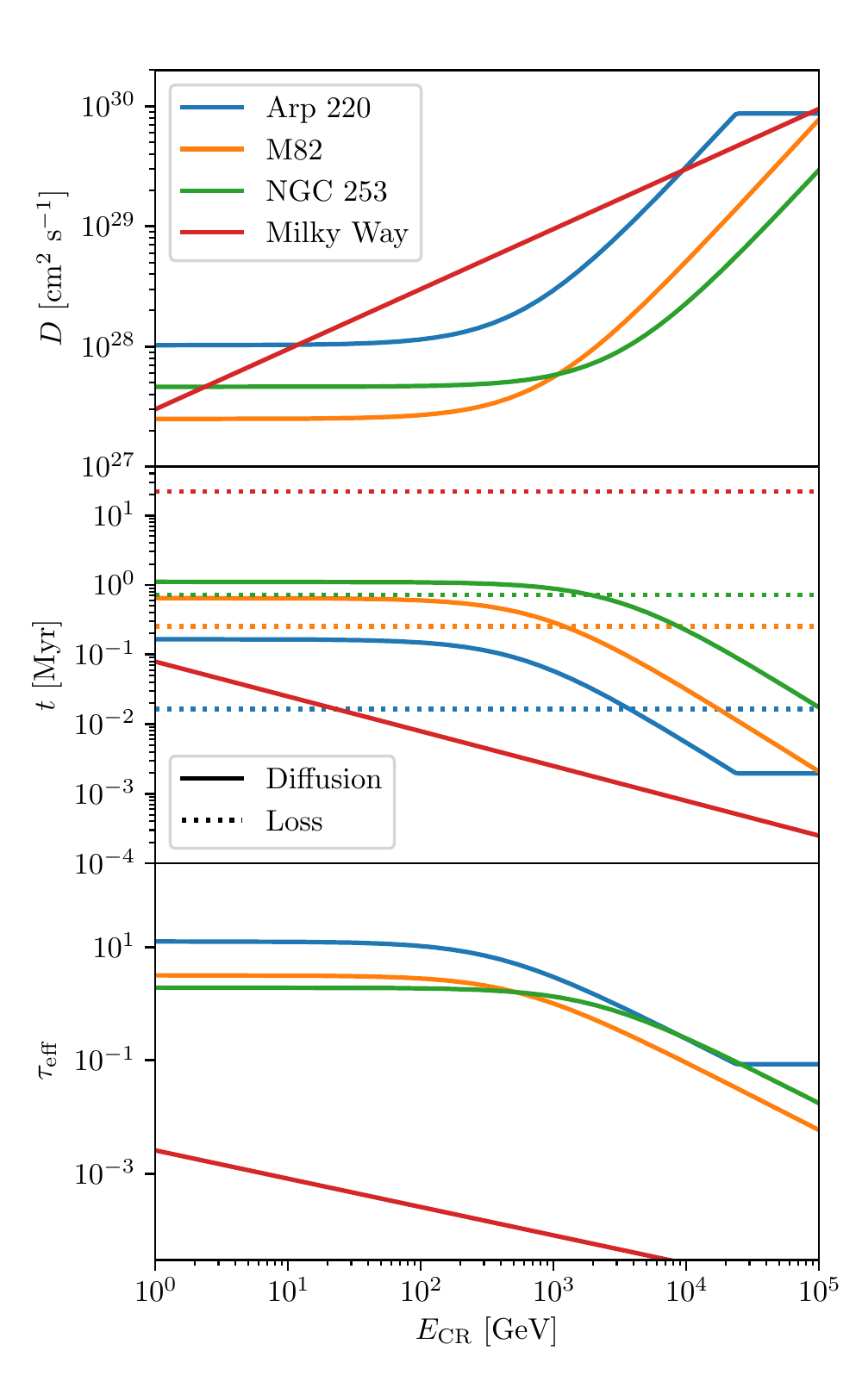}
    \caption{CR diffusion coefficients $D$ (top panel), diffusive escape and $pp$ loss timescales ($t_{\rm diff}$ and $t_{\rm loss}$, middle panel), and effective optical depth $\tau_{\rm eff}$ (bottom panel) as a function of CR energy. The lines shown for the starburst galaxies are computed using the fiducial parameters adopted for each of our starburst galaxies, as listed in \autoref{tab:starburst_properties}. The Milky Way lines use an empirical estimate of the diffusion coefficient (\autoref{eq:dcoef_mw}) together with the measured gas surface density and scale height (see \autoref{ssec:diff_model}). The kink in the lines for Arp 220 marks to the energy for which the CR streaming speed approaches $c$.}
    \label{fig:dcoef}
\end{figure}

We plot $D$ as a function of energy in the upper panel of \autoref{fig:dcoef}, using the full energy-dependent expression for $\vst$ (\autoref{eq:vratio}); this plot uses the fiducial parameters given in \autoref{tab:starburst_properties} for each galaxy, together with $\chi_{-4} = 1$ for all galaxies, $C_3 = 1$ for M82 and NGC 253, and $C_3 = 10$ for Arp 220. We also give numerical estimates of the total diffusion coefficient for our sample galaxies, assuming we are in the $\MA > 1$ regime, and at energies small enough that $\vst\approx \vAi$, in \autoref{tab:starburst_properties}. We see that diffusion coefficients are of order $10^{27}$ cm$^2$ s$^{-1}$ independent of energy, from $\approx 1 - 1000$ GeV. For comparison, we also plot an estimate of the diffusion coefficient for the Milky Way,
\begin{equation}
    D_{\rm MW}\approx 3\times 10^{27} \ECRz^{1/2}\;\mathrm{cm}^2\;\mathrm{s}^{-1}
    \label{eq:dcoef_mw}
\end{equation} 
the numerical value at 1 GeV is the empirical estimate from \citet{Gabici07a}, and the scaling $D\propto \ECR^{1/2}$ is that expected for a Kraichnan turbulent cascade \citep{Lacki13a}. Clearly the energy-dependence of $D$ is rather different in starbursts than in the Milky Way, even though the absolute value (at 1 GeV) is rather similar.

\section{Implications for $\gamma$-ray production}
\label{sec:gamma_rays}

Having derived an estimate for the CR diffusion rate in starburst galaxies, we next work out the implications for the observable $\gamma$-ray emission of these systems. In \autoref{ssec:diff_model} we develop a very simple model for the diffusive escape of CRs from a galaxy. Then in \autoref{ssec:calorimetry} we apply this model to compute the calorimetry of starburst systems, and in \autoref{ssec:spectra} we discuss the implications for their $\gamma$-ray spectra.

\subsection{Diffusion model}
\label{ssec:diff_model}

\subsubsection{Assumptions and equations}

Consider a galactic disc with total column density $\Sigma$ and scale height $h$; we will assume that the gas is distributed exponentially with height, so the density as a function of height $z$ is $\rho = \rho_0 e^{-z/h}$, where $\rho_0 = \Sigma/2h$. In general we expect $h\approx L$, as noted above. Suppose that a flux of cosmic rays is injected at $z=0$, and travels through the galactic disc before escaping. For simplicity we will make the following assumptions: (1) production or acceleration of CRs within the disc is negligible in comparison to those injected at the midplane; (2) $pp$ collisions are the only significant source of CR loss, and we neglect transfer of CRs between energies, i.e., once a CR loses its energy to a $pp$ collision, we assume it is removed from the population entirely; (3) there is no advective escape of CRs, only diffusive escape. This final assumption means that what we are computing is an upper limit on how calorimetric a galactic disc can be, since of course there may be advective escape. We return to this question below.

Under these assumptions, CR transport in the vertical direction obeys a simple diffusion equation with a loss term:
\begin{equation}
    \label{eq:diffeq}
    \frac{d}{dz}\left(-D \frac{d}{dz} U\right) = -\frac{U}{t_{\rm loss}},
\end{equation}
where $D$ is the diffusion coefficient, $U$ is the cosmic ray energy density at some chosen energy, and $t_{\rm loss}$ is the mean time required for a CR to be lost via a $pp$ collision. The term in parentheses on the left hand side is the CR flux, so this equation simply asserts that the change in flux through any height $z$ must be equal to the rate at which CRs are lost to $pp$ collisions at that height. Following our approximation that CRs are removed by a single interaction and do not transfer from one energy bin to another, this equation applies independently at each CR energy.

\subsubsection{Models for CR loss and diffusion}

In general the loss time and the diffusion coefficient can be functions of both the gas density and the local CR energy density. The dependence of the loss time is simple, however: for relativistic CRs, to good approximation we have \citep[e.g.,][]{Kafexhiu14a}
\begin{equation}
    t_{\rm loss} \approx \frac{1}{n \sigma_{\rm pp} \eta_{\rm pp} c},
\end{equation}
independent of energy. Here $\sigma_{\rm pp} \simeq 40$ mbarn\footnote{1 mbarn = $10^{-27}$ cm$^2$.} is the total hadronic collision cross section, $\eta_{\rm pp}\simeq 1/2$ is the elasticity, $n = \rho/\mu_p \mH$ is the number density of nucleons, and $\mu_p\approx 1.17$ is the number of nucleons per proton for a gas that is 90\% H and 10\% He by mass. 
The variation of the diffusion coefficient is somewhat more subtle, since in general it depends on the turbulent velocity dispersion, the Alfv\'en Mach number, and the ionisation fraction (\autoref{eq:dpar}). If the velocity dispersion is independent of height, and the Mach number that is relevant to the correlation length of the turbulence is that which prevails in the midplane (and thus is also independent of height), then the diffusion coefficient is simply proportional to the local ion Alfv\'en velocity $\vAi$. This in turn depends on the local density and ionisation fraction as $\vAi\propto (n \chi)^{-1/2}$. To first order, equating ionisation and recombination rates implies $\chi\propto (\zeta/n)^{1/2}$, and thus we expect a generic dependence $D\propto \vAi\propto (\zeta n)^{-1/4}$. More generally we can parameterise the dependence of $D$ on gas density and on CR energy density (to which we assume the primary ionisation rate $\zeta$ is proportional) as $D\propto n^{-\beta} U^{-\eta}$.

In order to render the problem analytically tractable, we will neglect variation in ionisation rate with height (i.e., we take $\eta=0$), on the grounds that the gas density varies much more strongly than the CR energy density over the bulk of the disc. We therefore adopt as a fiducial assumption that the diffusion coefficient varies as a function of the local gas density alone.\footnote{It is also worth noting that $\eta=0$ is also exactly what one expects for CR diffusion due to externally-driven turbulence in a fully ionised medium, with $\beta=1/4$ and $1/6$ corresponding to Kraichnan and Kolmogorov turbulent spectra, respectively  \citep[e.g.,][]{Lacki13a}. Thus the solutions we derive here are equally applicable to the case of CR transport by diffusion in ionised gas.} We provide solutions for arbitrary $\beta>0$, and adopt $\beta=1/4$ as our fiducial choice for all plots and numerical evaluations.\footnote{The condition $\beta > 0$ is required for the problem to be well-defined. If $\beta=0$, then one can show that the only solution to \autoref{eq:diffeq} is one in which the CR flux reaching infinity is identically zero, and if $\beta<0$ then no solution exists.} This simple powerlaw dependence cannot strictly hold as one moves out of the midplane into the corona, since at sufficiently low density the material must become ionised enough for the turbulent cascade to reach the CR gyroradius, at which point the cascade rather than the streaming instability will begin to provide the turbulence off which the CRs scatter. However, we show in \aref{app:diff_numerical} that the degree of calorimetry we compute below is only weakly sensitive to $\beta$, indicating that the precise functional form of $D$ has relatively little impact on the qualitative results. We similarly show that including a dependence of $D$ on the local CR energy density ($\eta \neq 0$) also leads to only small changes in the final result.

Inserting our expressions for $t_{\rm loss}$ and $D$ into \autoref{eq:diffeq} and non-dimensionalising, we have
\begin{equation}
    \label{eq:diffeq_nondim}
    \frac{d}{dx}\left(e^{\beta x} \frac{du}{dx}\right) = \tau_{\rm eff} e^{-x} u,
\end{equation}
where $x=z/h$, $u = U/U(0)$ and
\begin{eqnarray}
    \tau_{\rm eff} & = & \frac{\sigma_{\rm pp} \eta_{\rm pp} \Sigma h c}{2 D_0 \mu_p \mH}
    \nonumber \\
    & = & 9.9 \,\Sigma_3 h_2 D_{0,27}^{-1}
    \nonumber \\
    & = & 0.32 \, \Sigma_3 \uLAone^{-1} \chi_{-4}^{1/2} \max\left(\MA,\MA^4\right),
    \label{eq:taueff}
\end{eqnarray}
where $D_{0,27} = D_0/10^{27}$ cm$^2$ s$^{-1}$, $\Sigma_3 = \Sigma/10^3$ $M_\odot$ pc$^{-2}$, and $h_2=h/100$ pc; in the second line of numerical evaluation, we have used \autoref{eq:dpar} to substitute for $D$, and adopted $h=L$. The physical meaning of $\tau_{\rm eff}$ is straightforward: the quantity $\sigma_{\rm pp} \eta_{\rm pp} \Sigma/\mu_p\mH$ is the effective ``optical depth'' (column density times cross section) of the galactic plane to cosmic rays traveling in a straight line at $c$ perpendicular to the plane, while the quantity $h c/D_0$ is the ratio of the distance that a CR travels when diffusing a distance $h$ to the linear distance travelled; thus $\tau_{\rm eff}$ is the effective optical depth of the material through which a CR must pass on its way out of a galactic disc. We can also understand this relationship in terms of timescales, by defining $t_{\rm diff} = h^2/D_0$ as the characteristic time required for a CR to diffuse out of the disc, and $t_{\rm loss}$ evaluated with the mid-plane value of $n$ as the characteristic loss time; with these definitions, $\tau_{\rm eff}$ is simply the ratio of $t_{\rm loss}$ to $t_{\rm diff}$.

We plot $t_{\rm diff}$ and $t_{\rm loss}$, and $\tau_{\rm eff}$, as a function of $\ECR$ for our fiducial starburst properties in the lower two panels of \autoref{fig:dcoef}. Clearly the starbursts will have $\tau_{\rm eff} > 1$ out to energies of at least $\approx 1$ TeV. For comparison we show the Milky Way escape and loss times; for this purpose we adopt a gas surface density of $\Sigma\approx 14$ $M_\odot$ pc$^{-2}$ \citep{McKee15a} and a scale height of $h\approx 140$ pc \citep{Boulares90a}. Clearly the Milky Way is the opposite limit from the starbursts at nearly any CR energy, and has $\tau_{\rm eff} \ll 1$.

\subsubsection{Boundary conditions}

To solve \autoref{eq:diffeq_nondim} we require two boundary conditions. One is simply that $u(0) = 1$ by construction, but of course this does not fix the scaling between $u$ and $U$. To do so, we require that the flux at the midplane match the value at which CRs are being injected into the galaxy. Let the star formation rate per unit area be $\dot{\Sigma}_*$, let $M_{\rm SN}$ be the mass of stars that must be formed per SN, and let $E_{\rm CR,SN}$ be the energy injected into the CR population per SN. In this case, the boundary condition at the midplane can be expressed as
\begin{equation}
    -\left(D\frac{d}{dz} U\right)_{z=0} = \frac{1}{2} E_{\rm CR,SN} \frac{\dot{\Sigma}_*}{M_{\rm SN}}.
\end{equation}
Note that the factor of $1/2$ is because half the CRs escape into the upper half-plane at $z>0$, and the other half into the lower half-plane at $z<0$. This in turn requires that
\begin{equation}
    U(0) = -\left(\frac{1}{u'(0)}\right)\frac{h E_{\rm CR,SN} \dot{\Sigma}_*}{2 D_0 M_{\rm SN}} \equiv -\frac{\Phi_0}{u'(0)},
\end{equation}
where the prime indicates differentiation with respect to $x$, and
\begin{eqnarray}
    \Phi_0
    & = & 5.1\times 10^{-8} \,h_2 \dot{\Sigma}_{*,2} D_{0,27}^{-1}\;\mathrm{erg}\,\mathrm{cm}^{-3}
    \nonumber \\
    & = & 1.7\times 10^{-9}\;\mathrm{erg}\,\mathrm{cm}^{-3} \cdot {}
    \nonumber \\
    & & \qquad \frac{\dot{\Sigma}_{*,2}\chi_{-4}^{1/2}}{\uLAone}  \max\left(\MA,\MA^4\right),
    \label{eq:phi0}
\end{eqnarray}
where $\Phi_0$ is the CR flux at $z=0$ measured in units where the scale height is unity and $\dot{\Sigma}_{*,2} = \dot{\Sigma}_*/100$ $M_\odot$ pc$^{-2}$ Myr$^{-1}$. Our numerical evaluation here is for $M_{\rm SN}=100$ M$\odot$ and $E_{\rm CR,SN}=10^{50}$ erg. Note that the value of $\Phi_0$ affects only the normalisation of the CR energy density versus height, not its shape.

The second boundary condition applies at the top of the disc. As a fiducial choice, we note that, at sufficiently large height and low density, the rate of CR transport should approach the streaming speed, since the CR mean free path is proportional to $e^{\beta z/h}$ and therefore at large enough $z$ becomes larger than the height; at this point the CRs should simply free-stream to infinity. Following \citet{Zweibel17a} we therefore require that the CR flux, $-D (dU/dz)$, approach $\vst (U + P) = (4/3) \vst U$ as $z\to\infty$, where $P = U/3$ is the CR pressure. This condition is satisfied if
\begin{eqnarray}
    -\left(e^{\beta x} \frac{d\ln u}{dx}\right)_{x\to\infty} & = & \frac{4h \vst}{3 D_0} \equiv \Upsilon_0
    \nonumber \\
    & = & 41.2\, h_2 V_{\rm St,3} D_{0,27}^{-1}
    \nonumber \\
    & = & \frac{4}{3} \MA^3
    \label{eq:upsilon}
\end{eqnarray}
where $\Upsilon_0$ is the dimensionless CR streaming velocity as $z\to\infty$, $V_{\rm St,3} = \vst/10^3$ km s$^{-1}$, and in the final line we have used \autoref{eq:vstream} and \autoref{eq:dpar} to evaluate $\vst$ and $D$. This provides our second boundary condition. For the solutions we obtain below and our fiducial parameter choices, the transition to free-streaming generally occurs at $x\sim 20$, corresponding to $\sim 2$ kpc. We discuss the outer boundary condition further in \aref{app:diff_numerical}, where we show that our qualitative results are quite insensitive to any reasonable choice.

\subsubsection{Solutions}

The solution to \autoref{eq:diffeq_nondim} for any $\beta>0$ is
\begin{equation}
    u = e^{-\beta x/2} \frac{a I_{\beta/(\beta+1)}\left(d e^{-(\beta+1)x/2}\right) + b I_{-\beta/(\beta+1)}\left(d e^{-(\beta+1)x/2}\right)}{a I_{\beta/(\beta+1)}\left(d\right) + b I_{-\beta/(\beta+1)}\left(d\right)},
    \label{eq:Usol}
\end{equation}
where
\begin{eqnarray}
a & = & \Gamma\left(\frac{\beta}{\beta+1}\right) \Upsilon_0 \\
b & = & \frac{\left(\beta+1\right)^{2/(\beta+1)}}{\beta} \Gamma\left(\frac{1}{\beta+1}\right) \tau_{\rm eff}^{\beta/(\beta+1)} \\
d & = & \frac{2}{\beta+1}\sqrt{\tau_{\rm eff}},
\end{eqnarray}
$I_n(x)$ is the modified Bessel function of the first kind of order $n$, and $\Gamma(x)$ is the usual $\Gamma$ function.

We evaluate $\tau_{\rm eff}$, $\Phi_0$, $\Upsilon_0$, and the midplane CR energy density $U(0) = -\Phi_0/u'(0)$ for $\beta=1/4$ and the fiducial parameters for each of our galaxies in \autoref{tab:starburst_properties}, using $E_{\rm CR,SN} = 10^{50}$ erg and $M_{\rm SN} = 100$ $M_\odot$, i.e., assuming 1 supernova per 100 $M_\odot$ of stars formed, with that supernova putting $\approx 10\%$ of its energy into CRs. For our fiducial parameters, CR midplane energy densities are of order $1-5$ keV cm$^{-3}$, similar to  previously-published estimates  \citep[e.g.,][]{Yoast-Hull16a}; however, we remind readers that we have explicitly assumed that advective escape of CRs is negligible, so if advective escape is significant the true energy densities may be somewhat lower. These energy densities are roughly a factor of $10^3-10^4$ larger than in the Milky Way. If this ratio also applies to the lower energy CRs that dominate ionisation, these values are consistent with our fiducial choices of CR ionisation rates in \autoref{ssec:starburst_ISM}, and thus our fiducial estimate of $\chi$.

\subsection{Calorimetry}
\label{ssec:calorimetry}

We can immediately use our solution for the diffusion model to derive an analytic expression for how calorimetric a particular galaxy is, i.e., what fraction of the injected CR flux will be lost to $pp$ collisions, and what fraction will escape to infinity. The dimensionless flux at any height is $-e^{\beta x} u'$, and so the ratio of the flux reaching infinity to the flux injected at the midplane is
\begin{equation}
    f_{\rm esc} = \lim_{x\to\infty} e^{\beta x} \frac{u'(x)}{u'(0)}
\end{equation}
The calorimetric fraction, which we define as the fraction of the injected CRs that are lost to $pp$ collisions and thus produce $\gamma$ rays, is simply $f_{\rm cal} = 1-f_{\rm esc}$. Inserting our analytic solution for $u$ (\autoref{eq:Usol}) and doing some algebraic simplification, we obtain
\begin{eqnarray}
    f_{\rm cal} & = & 1 - \left[ {}_0F_1\left(;\frac{\beta}{\beta+1};\frac{\tau_{\rm eff}}{\left(\beta+1\right)^2}\right) +
    {}
    \right.
    \nonumber \\
    & & \qquad \left.
    \frac{\tau_{\rm eff}}{\Upsilon_0} {}_0F_1\left(;\frac{\beta+2}{\beta+1};\frac{\tau_{\rm eff}}{\left(\beta+1\right)^2}\right)\right]^{-1},
    \label{eq:fcal}
\end{eqnarray}
where $_{0}F_1(;a;x)$ is a generalized hypergeometric function.\footnote{The hypergeometric function $_{0}F_1(;a;x)$ is related to the more commonly-encountered Bessel function by $_{0}F_1(;a;x) = \Gamma(a) x^{(1-a)/2} I_{a-1}(2\sqrt{x})$.} Note that the value of $\Upsilon_0$ has only a tiny effect on $f_{\rm cal}$ unless $\Upsilon_0 \ll 1$; thus to first order $f_{\rm cal}$ depends only on $\tau_{\rm eff}$. As pointed out above, the value of $\Phi_0$ has no effect on $f_{\rm cal}$ at all, and merely serves to set the midplane energy density.

\begin{figure}
    \centering
    \includegraphics[width=\columnwidth]{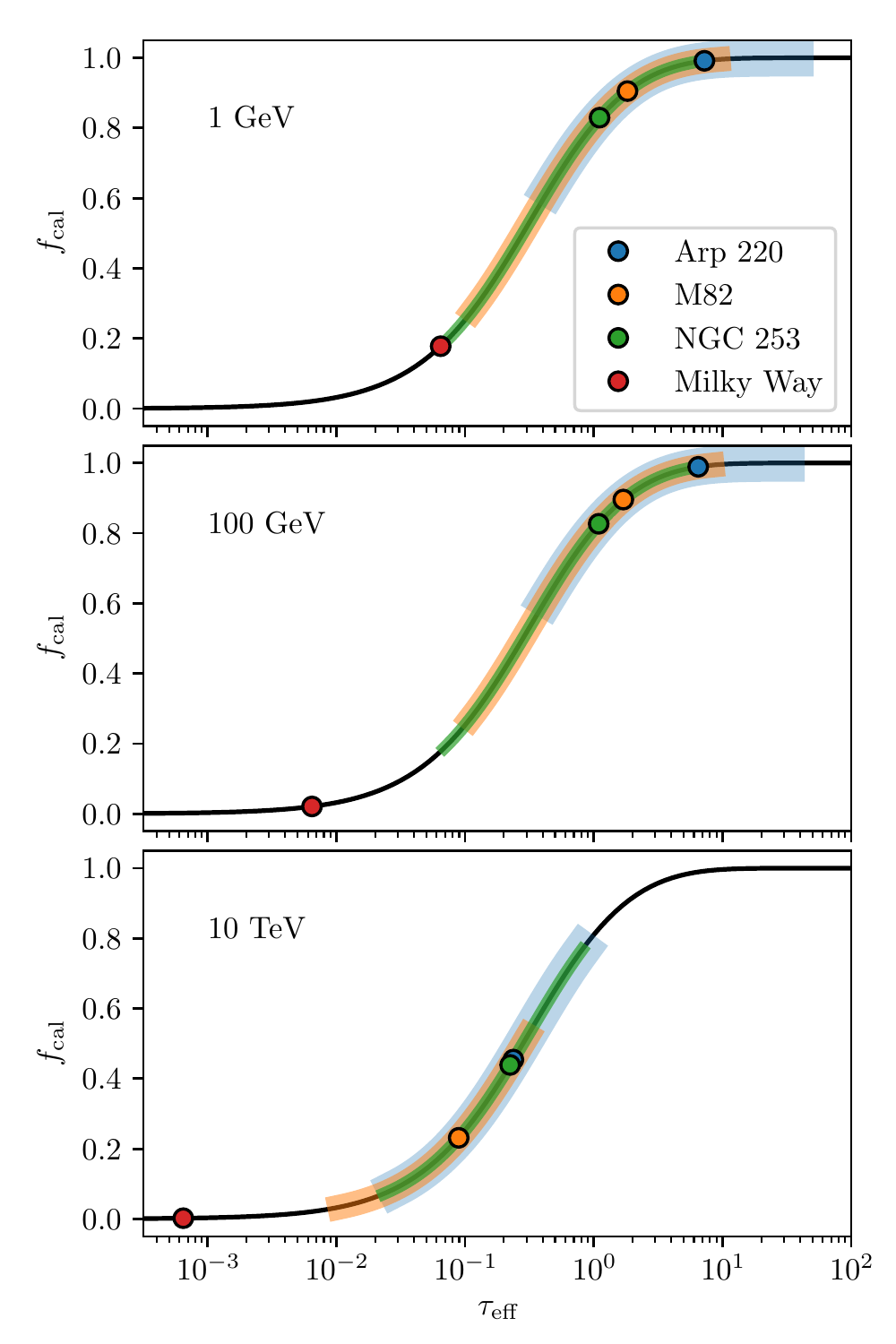}
    \caption{Calorimetric fraction $f_{\rm cal}$ as a function of effective optical depth $\tau_{\rm eff}$ for our fiducial parameters $\beta=1/4$ and $\Upsilon_0=10.7$. The black line, which is the same in every panel, is \autoref{eq:fcal}. Coloured points mark our fiducial estimates of $\tau_{\rm eff}$ and $f_{\rm cal}$ for Arp 220, M82, NGC 253, and the Milky Way Solar neighborhood, at CR energies of 1 GeV, 100 GeV, and 10 TeV as indicated in the panels. For Arp 220, M82, and NGC 253, the coloured bands indicate the range of variation in produced by varying $\MA$ over the range $1-3$.}
    \label{fig:fcal}
\end{figure}

We plot the general relationship between $f_{\rm cal}$ and $\tau_{\rm eff}$ in \autoref{fig:fcal}, and in each of the panels of that figure we also show estimates of $\tau_{\rm eff}$ and thus $f_{\rm cal}$ for our sample starburst galaxies at three different CR energies, using the fiducial values given in \autoref{tab:starburst_properties}. In this plot, coloured bands show a plausible uncertainty range about our central estimate, corresponding to adopting Alfv\'en Mach numbers $\MA=1-3$, compared to our fiducial value $\MA=2$. Consistent with what is shown in \autoref{fig:dcoef}, the calorimetric fraction for our sample starbursts is nearly independent of energy for energies $\lesssim 1$ TeV, and we give numerical values for $f_{\rm cal}$ over this energy range in \autoref{tab:starburst_properties}. Over this energy range, the starbursts have calorimetric fractions of $\approx 80-100\%$, with the plausible range extending down to $f_{\rm cal}\approx 30\%$ for NGC 253 and M82. Only at CR energies $\gtrsim 10$ TeV does the calorimetric fraction begin to decline. However, we do remind readers that the calorimetric fractions we have computed are only upper limits, because they apply only to CRs that are deposited into the neutral phase of the ISM. CRs that are trapped inside a supernova remnant that escapes from the galactic disc in a chimney may advect out along with the escaping gas, thereby never interacting with the bulk of the ISM at all; we defer further discussion of this possibility to \autoref{ssec:advection}.

For comparison with the starbursts, we also plot estimates of $\tau_{\rm eff}$ and $f_{\rm cal}$ for the Solar neighbourhood in the Milky Way using the same model. For this purpose we adopt the same empirical diffusion coefficient given in \autoref{eq:dcoef_mw}, and we compute $\tau_{\rm eff}$ using the same numerical values shown in \autoref{fig:dcoef}.\footnote{As noted above, our diffusion model with $\beta=1/4$ is applicable to the Milky Way if we assume the turbulence has a Kraichnan spectrum, and the CR energy densities we derive under this assumption are reasonable: at 1 GeV, \autoref{eq:taueff} gives $\tau_{\rm eff} \approx 0.065$, and the Solar neighbourhood's observed star formation rate is $\dot{\Sigma}_*\approx 2.5\times 10^{-5}$ $M_\odot$ pc$^{-2}$ Myr$^{-1}$ \citep{Fuchs09a}, corresponding to  $\Phi_0\approx 1.2\times 10^{-12}$ erg cm$^{-3}$ (\autoref{eq:phi0}). Using these figures in \autoref{eq:Usol} implies a midplane cosmic ray energy density of $U(0) \approx 2.8$ eV cm$^{-3}$, in reasonable agreement with empirical estimates \citep[e.g.,][]{Drai11}.} We see that the calorimetric fraction of the Milky Way is small at 1 GeV ($f_{\rm cal} \approx 0.18$), and drops precipitously at higher energies.

\subsection{$\gamma$-ray spectra}
\label{ssec:spectra}

We next calculate the $\gamma$-ray spectra of our three representative starbursts; these have all been detected in the high-energy  \citep[$\sim$GeV;][and references therein]{Peng16a,Griffin16a,Fermi19a} and -- in the case of M82 \citep{VERITAS09a} and NGC253 \citep{HESS18a} -- very high energy ($\sim$TeV) $\gamma$-ray regimes and, as mentioned above, considerable theoretical and modelling effort has been devoted to understanding their
$\gamma$-ray emission. 

In our modelling, we incorporate the impact of the energy-dependent calorimetry fractions calculated above on the {\it in situ}, steady state cosmic ray proton distributions (we ignore the impact of heavier ions in the CR beam for simplicity) and the resulting hadronic $\gamma$-ray emission of each system. 
For the purposes of our calculation, we assume that the underlying parent proton distribution (before escape) is a pure power-law in momentum with an abrupt cut-off at $10^{15}$ eV.
We calculate hadronic $\gamma$-ray emission following the prescription of \citet{Kafexhiu14a}.
This calculation is not entirely without tunable parameters.
In particular, it is to be emphasised that our model does not prescribe the  spectral index of the cosmic ray proton distribution at injection into a galaxy's ISM nor the absolute normalisation of the $\gamma$-ray luminosity from the entire galaxy.
In a rigorous phenomenological treatment, 
best fit values of these would be obtained through a fitting procedure. 
However,
as our purpose here is the  demonstration that the FLRW scenario can reproduce the qualitative features of the starbursts' $\gamma$-ray emission, we do not present such fitting here.
Rather we adopt ``off-the-shelf" spectral indices for the star-bursts nominated in the {\it Fermi} 8-year catalogue \citep[][viz. NGC 253:  $2.14 \pm 0.050$,
M82:  $2.20 \pm 0.041$,
Arp 220:  $2.42 \pm 0.13$]{Fermi19a}\footnote{Note the interesting point that the $\sim$GeV spectral index of Arp 220 is incompatible at the $\sim 2 \sigma$ level with the indices of the two more moderate starbursts.
In our model, this should correspond to a mirroring difference in the spectral indices of the CR proton populations {\it at injection} into the ISM for these galaxies.
As currently conceived, our model does not explain this difference. However, we also note that it is possible that the lowest energy bins for Arp 220 contain some contribution from either contamination by a nearby source, or inverse Compton or bremsstrahlung emission from CR electrons, which are also not included in our model.};
and, for the absolute normalization of the $\gamma$-ray flux, we fit by eye to find the efficiency of cosmic ray acceleration per core-collapse supernova required to obtain consistency with the observations.
Pleasingly, when assuming a standard total mechanical energy release of 
$10^{51}$ erg per core collapse supernova,
 this exercise results in cosmic ray acceleration efficiencies of $\sim$ 10\%\footnote{In other words, $10^{50}$ erg in CR protons per core-collapse supernova} in Arp 220 and NGC 253, while M82 requires $\sim $5\%; these values are entirely consistent with estimates obtained from observation of many different environments, including the Milky Way \citep[e.g.][]{Strong10a}.
 
 A number of other inputs are required for our calculation: The star-formation rates  we adopt are 220 $M_\odot$ yr$^{-1}$ for Arp 220
 \citep{Varenius16a},
 4.0 $M_\odot$ yr$^{-1}$ for M82 \citep{Strickland09a}, and 
1.7  $M_\odot$ yr$^{-1}$ for NGC 253 \citep{Bendo15a}.
Distances to the systems, starburst radii, and energy densities in the various interstellar radiation fields we adopt from table 3 of \citet{Peretti19a}.
The latter two are necessary for us to calculate the $\gamma \gamma \to$ e$^\pm$ opacity of the starburst systems, which we do following the prescription in Appendix D of \citet{Peretti19a}.

\begin{figure}
    \centering
    \includegraphics[width=\columnwidth]{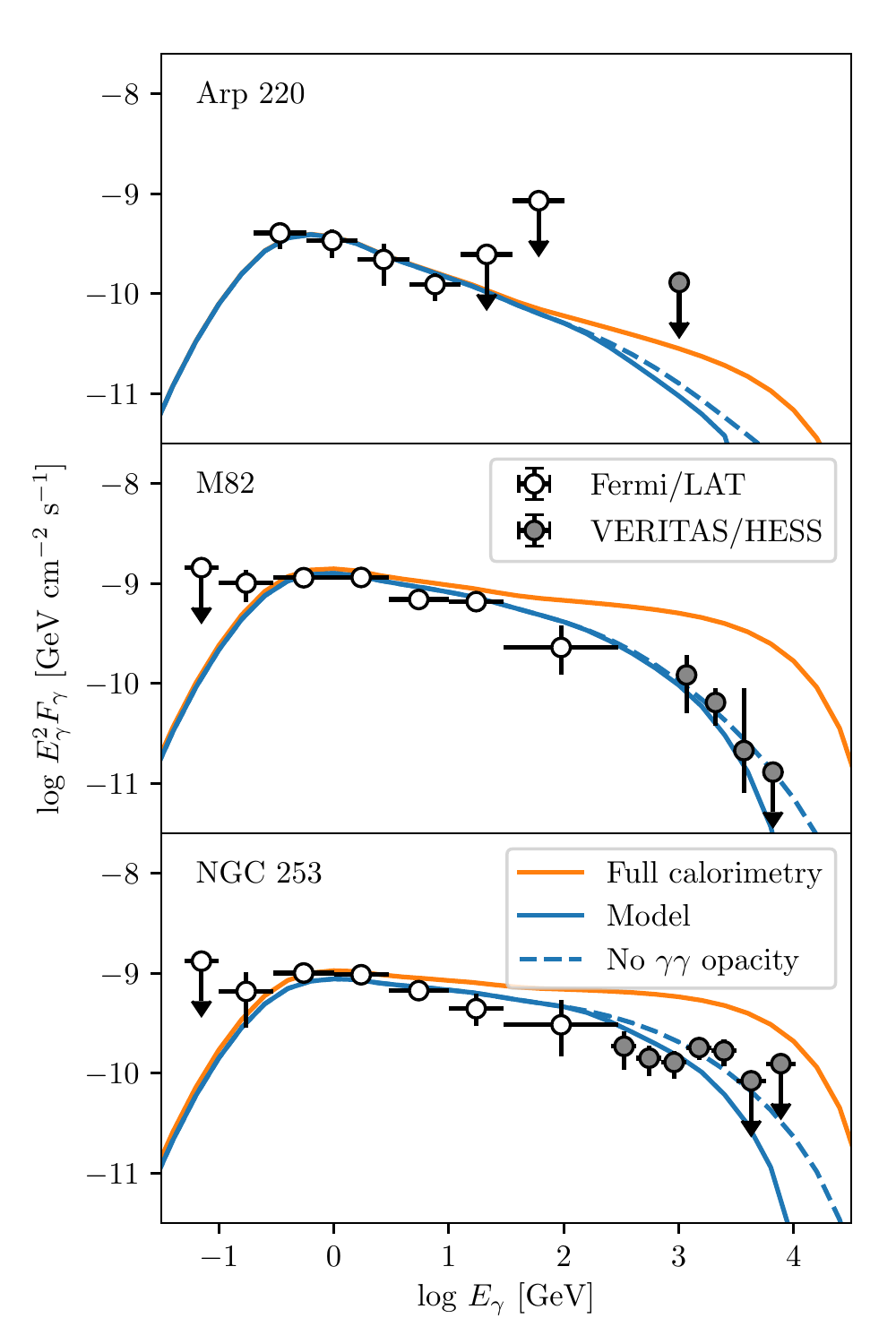}
    \caption{Spectra of NGC 253, M82, and Arp 220 computed as described in the main text. Solid blue lines show the standard model, dashed blue lines show the predictions if we ignore the effects of $\gamma\gamma$ opacity, and orange lines show the spectra that would be expected for perfect calorimetry independent of CR energy.  Circles with error bars show observations. Open circles are from the \textit{Fermi}/LAT 8-year catalog \citep{Fermi19a} for NGC 253 and M82, and from the analysis of \textit{Fermi} data by \citet{Peng16a} for Arp 220. Grey filled circles are from VERITAS for M82 \citep{VERITAS09a} and Arp 220 \citep{Fleischhack15a}, and from HESS for NGC 253 \citep{HESS18a}. Horizontal error bars show the energy band over which a particular observation is made, while vertical error bars show $1\sigma$ uncertainties; points with downward arrows indicate $1\sigma$ upper limits for bands without detections.}
    \label{fig:grspectra}
\end{figure}

We display the results of this exercise in \autoref{fig:grspectra}.
The onset of energy-dependent escape can be seen to affect the theoretical $\gamma$-ray curves for energies $\gtrsim$100 GeV\footnote{As seen for orange ``full calorimetry" curves, this is well below the energy where the assumed $10^{15}$ eV cut-off of the parent proton distribution starts to affect the $\gamma$-ray spectrum.}.
Unfortunately, $\gamma \gamma$ pair production opacity starts to have an impact in these systems at roughly similar energies.
Nevertheless, one can see that the energy-dependent escape we have determined has the larger effect and, for the TeV-detected M82 and NGC 253, it is this effect that contributes most to the rather good matches to the broadband data. While the spectral softening at TeV energies required by the data could also in principle be reproduced by a turn-down in the injected CR spectrum, there is no physically compelling reason to suppose that such a turn-down exists, and the energy-dependent diffusion coefficient we have computed from first principles reproduces the feature without the need for any such fine-tuning.
%
Perhaps most importantly, the fact that 
our FLRW mechanism produces an energy-independent calorimetry fraction for CRs with energies $\lesssim$ TeV means that the starburst spectra are hard up to $\sim 100$ GeV energies, matching the observations.

\section{Discussion}
\label{sec:discussion}

Now that we have shown that our calculation of CR transport in starbursts produces results in reasonable agreement with observations, in this section we proceed to work out some additional implications of our calculation.

\subsection{Bulk turbulent mixing of CRs is unimportant compared to diffusion}

We have argued in this paper that the dominant mechanism for CR transport in starburst galaxies is FLRW, with the speed of streaming along field lines set by the balance between streaming instability and CR damping. It is interesting to compare this process to other possible transport mechanisms. Since we have found that CRs stream only at the Alfv\'en speed, one immediate question is whether they might be transported just as effectively simply by bulk motion of the gas. However, we can dispense with this possibility rather quickly. If we imagine that CRs were completely frozen into the gas and were transported simply by the turbulent mixing of that gas, the natural diffusion coefficient would be
\begin{equation}
    D_{\rm mix} \approx \frac{1}{3} \sigma L.
\end{equation}
Comparing this expression to \autoref{eq:dpar}, we see that
\begin{equation}
    \frac{D_{\parallel}}{D_{\rm turb}} \approx \frac{3\vAi\LA}{\sigma L}
    \approx \sqrt{\frac{3}{\chi}} \min\left(\MA^{-1},\MA^{-4}\right).
\end{equation}
Thus we have $D_{\rm mix} < D_{\parallel}$ as long as
\begin{equation}
    \MA \lesssim \left(\frac{3}{\chi}\right)^{1/8} = 3.6 \chi_{-4}^{-1/8}.
\end{equation}
Thus unless the Mach number of the turbulence is significantly greater than suggested by dynamo theory, or the ionisation fraction is much higher than predicted by our chemical models ($\sim 10^{-2}$ rather than $\sim 10^{-4}$), we expect streaming-limited diffusion plus FLRW to dominate over turbulent mixing as a transport process.

\subsection{Diffusion versus advective loss}
\label{ssec:advection}

Advection of CRs directly into the wind escaping from starbursts may also play a role in determining their  $\gamma$-ray spectrum and luminosity. 
Scenarios invoking advection in this context  \citep[e.g.,][]{Yoast-Hull13a,Peretti19a} have a number of points in their favour: starbursts do generally drive global winds \citep[e.g.,][]{Veilleux05a} whose speeds -- in the 100s or even 1000s of km s$^{-1}$ range -- are large enough that advective escape can, {\it prima facie}, often compete with losses. Moreover, as for the ion-neutral-damping-limited streaming discussed above, advective escape is an energy-independent process that can (over the CR energy range where advection is dominant over diffusion) explain the phenomenologically required hard $\gamma$-ray spectrum found in starburst systems. 


However, advective escape encounters a significant problem, which our model avoids. The midplane of a starburst is dominated by neutral material, most of which is not in fact flowing out of the galaxy in a wind -- even generous estimates of the mass loading of the neutral material imply a gas consumption time that is many orbital periods long \citep[e.g.][]{Veilleux05a}. Indeed, it is unclear if any cold, neutral gas is entrained into the wind at all, or if neutral material only re-condenses out of the hot phase at some distance from the host galaxy \citep[e.g.,][]{Thompson16b, Krumholz17e, Schneider17a}. Thus if CRs are to escape via advection, they must be advecting mainly with the hot, ionised medium, not the cold, neutral one.

This is problematic from the standpoint of explaining galaxy spectra, however. While the column density of the neutral material is sufficient to render CRs calorimetric, this is not the case for the hot gas, which has a much smaller total column. Thus it is not clear how one can simultaneously have advective escape (which must mean CRs are in the hot phase) and yet still produce the hard $\gamma$-ray spectra that are required by observations.\footnote{A similar problem exists for the radio emission produced by CR electrons -- the observed hard radio spectra of starbursts require that CR electrons undergo  bremsstrahlung or ionisation losses at a rate that is at least comparable to their rate of inverse Compton losses, but this would only be expected if those CRs were interacting primarily with dense, cold, neutral gas, rather that diffuse, hot, and ionised material -- see \citet{Lacki10a} and \citet{Thompson13a} for further discussion.} Existing models for starburst $\gamma$-ray spectra mostly approach this problem by leaving the advective escape time and $pp$ loss times as free parameters \citep[e.g.,][]{Yoast-Hull13a, Peretti19a}. Fitting the spectra requires that both timescales be $\sim 1$ Myr. This is a reasonable advection timescale for the escape of hot gas from a starburst, but not for the escape of cold gas; conversely, this is a reasonable loss time for CRs moving through dense, neutral medium, but not for CRs encountering hot and diffuse gas. Thus having both the advective loss and $pp$ loss times near $\sim 1$ Myr requires a degree of fine-tuning: the CRs must spend enough time in the neutral phase to undergo significant losses, but must also then find their way into the hot phase in order to escape on the same timescale.

Our model avoids the need for such fine-tuning. We assume that CRs are injected into the cold phase that dominates the volume at the midplane, and thus we naturally recover a $pp$ loss time of $\sim 1$ Myr, as required. However, we also naturally find $\sim 1$ Myr as the energy-independent CR escape time (c.f.~\autoref{fig:dcoef}), not because CRs advect outward, but thanks to FLRW plus streaming-limited diffusion. Any CRs that are injected into the hot phase (which our model does not require but also does not forbid) will have $pp$ loss times much larger than their advective escape times, and will produce few $\gamma$-rays. This will reduce the degree of calorimetry, but not alter the hard spectrum.

A second point in favour of our model seems to be that -- for the two starbursts for which we have both high energy and very high energy data (M82 and NGC 253) -- our scenario provides a rather natural way to both understand the hard GeV spectra of the starbursts and to reconcile the GeV and TeV spectra in a fashion that  requires neither fine-tuning of a putative parent proton distribution cut-off nor uncomfortably steep CR injection spectra.
Indeed, in calculating the broadband $\gamma$-ray emission of our systems we adopted off-the-shelf spectral indices obtained from fitting to the GeV data alone, and yet, the spectra we calculate are evidently a rather good match at TeV energies.
In the work of, e.g., \citet{Peretti19a} or \citet{Wang18a}, in contrast, GeV and TeV data are fitted simultaneously for the M82 and NGC 253 cases and the resulting spectral indices are somewhat steeper than what we seem to need.

\subsection{Cosmic rays pervade the neutral gas of starbursts}

CRs undergo rapid losses in the dense, neutral environments found in starbursts, and we find that that these galaxies are nearly calorimetric. This raises the question of whether CRs should be able to spread out through all the neutral gas in a starburst galaxy at all, or whether they will be confined primarily to a small volume of the galaxy. It is therefore of interest to compute the CR filling factor $Q_\text{CR}$, which \citet{Lacki13a} defines as the overlap fraction of the total volume of all CR `bubbles' around individual SNRs -- which act as short-lived accelerators -- and the starburst region total volume.
Such CR bubbles will have radii $R_\text{RM} \sim \sqrt{D t_\text{CR}}$ where $t_\text{CR}$ is the minimum of the pertinent loss or escape times. \citeauthor{Lacki13a} shows that
\begin{equation}
    Q_{\rm CR} = \frac{8 \pi}{ 15} 
    \dot{\rho}_{\rm SN} D^{3/2} t_{\rm CR}^{5/2}
\end{equation}
where $\dot{\rho}_{\rm SN}$ is the volumetric injection rate density of SNe (i.e., SNe / unit volume / time). 
Setting $t_{\rm CR} = t_{\rm loss}$, we have
\begin{equation}
\dot{\rho}_{\rm SN} = \frac{\dot{\Sigma}_\star}{M_{\rm SN}  h} 
\end{equation}
and given typical values for $D$, the diffusion coefficient determined above, $h$, and $M_{\rm SN}\approx 100$ $M_\odot$, we find
$Q_{\rm CR} \sim 10 - 10^3 \gg 1$ (cf. \autoref{tab:starburst_properties}).

The large value of $Q_{\rm CR}$ implies that CRs do in fact fill all the neutral volume in starbursts. The essential reasons for this very large overlap are that, for the three star-burst systems, 
i)
$R_\text{CR} = \sqrt{D  t_\text{loss}} \sim 10s$ pc (see below) is large (but still only a fraction of $h$; see below) and ii) $\dot{\rho}_\text{SNR}$ is also large; within a volume of size $h^3$ over a time $t_\text{loss}$, the expected number of individual SNe is $\sim 10^5 - 10^6$ for our starbursts. Thus we have arrived at the interesting and seemingly paradoxical situation that, while we predict that the starbursts we investigate are $\gtrsim 50$\% calorimetric (i.e., $\lesssim$ 50\% of CR energy is lost outside the galaxies), at the same time CRs reach most of the neutral gas.
How does this happen?
The radii of the CR bubbles in Arp 220, M82, and NGC 253 we predict to be, respectively, 17, 34, and 75 pc, or, normalised to these systems' scale heights, about $h/4$, $h/2$, and $h/2$. 
Thus, to escape CRs, need to go another factor of $2 - 4$ in distance. 
However, the distance travelled only goes up as the square root of time in diffusive propagation, so the required time to escape is $\sim 16, \sim 4$, and $\sim 4 \times t_\text{loss}$ in the three systems. 
At the same time, CRs are being delivered essentially in situ all throughout the neutral gas because of the large density of active sources at any one time. 
These facts explain how the large calorimetry fraction we compute can be naturally reconciled with a large cosmic ray filling factor in starbursts.
Note that our findings here are different from that of \citet{Lacki13a} who, in his favoured scenario, posited that CR transport in neutral gas would be dominated by turbulent mixing and, on this basis, found $Q_\text{CR} \ll 1$ for this ISM phase.

\subsection{Why are starbursts mostly calorimetric?}

As a final topic of interest, we point out that our model provides a simple and clear explanation for why starbursts tend to be calorimetric with regard to the bulk of the CR population, and how the degree of calorimetry depends on the large-scale properties of the starburst. The degree of calorimetry is mostly controlled by the parameter $\tau_{\rm eff}$ (\autoref{eq:taueff}), which depends on the gas column density, the scale height, the diffusion coefficient, and various microphysical constants. We can gain some insight into the dependence of this parameter on the large-scale properties of galaxies by using our estimate for the macroscopic diffusion coefficient $D$, \autoref{eq:dpar}, in \autoref{eq:taueff}, and considering CRs near the peak of the energy distribution, where $\vst\approx \vAi$. Doing so, rewriting $\vAi = \vAt/\sqrt{\chi} \approx \sigma/\MA\sqrt{2\chi}$, and making use of \autoref{eq:Qgas} to substitute for $\sigma$, we obtain the general relationship
\begin{equation}
    \tau_{\rm eff} \approx \frac{\sqrt{2\chi} \sigma_{\rm pp} \eta_{\rm pp} c \MA^4}{\mu_{\rm p} \mH G Q_{\rm gas} t_{\rm orb}} \approx \chi_{-4}^{1/2} \mathcal{M}_{\rm A,2}^4 Q_{\rm gas,2}^{-1} \frac{16.5\;\mathrm{Myr}}{t_{\rm orb}},
    \label{eq:taueff_sb}
\end{equation}
where $\mathcal{M}_{\rm A,2}=\MA/2$ and $Q_{\rm gas,2}=Q_{\rm gas}/2$.

Our result in \autoref{eq:taueff_sb} is interesting in that it suggests that the degree of calorimetry depends, to first order, only on the galaxy rotation period. All other things being equal, galaxies with orbital periods $\gg 20$ Myr (essentially all non-starburst galaxies) will tend to be transparent to CRs and thus mostly non-calorimetric, while those with orbital periods $\ll 20$ Myr (essentially all starbursts) will tend to be opaque and thus achieve a high degree of calorimetry. This very general result follows from the interaction of three physical effects. The first is the turbulent dynamo, which tends to saturate at $\MA\approx 2$ and thus sets the strength of the magnetic field that inhibits CR streaming. The second is gravitational instability, which tends to saturate at $Q_{\rm gas}\approx 2$, and thus sets the combination of gas surface density and velocity dispersion to a value that depends only on the orbital period. The third is ionisation balance, which, at the high CR injection rates found in starbursts, tends to pick out a relatively high ionisation fraction $\chi\sim 10^{-4}$, even in molecular gas. The combination of these effects with the microphysical constants describing the interaction of CRs with the ISM serve to set the line between mostly calorimetric and mostly non-calorimetric galaxies at a galaxy orbital period of order 20 Myr.

\section{Summary and conclusion}
\label{sec:conclusions}

In this paper we present first-principles calculations of the mechanisms of CR transport in starburst galaxies. Such galaxies are distinguished by their very high densities, which result in their midplanes being dominated by neutral molecular material. We use a grid of astrochemical models to show that, despite the high CR ionisation rates expected in starbursts, typical ionisation fractions should be at most $\chi \approx 10^{-4}$. This is small enough that ion-neutral damping halts turbulent cascades at length scales far larger than the CR gyro-radius for CRs with energies up to hundreds of TeV. Consequently, these CRs cannot scatter off extrinsically-driven turbulence, and instead scatter off turbulence that is self-generated by the streaming instability. We show that the competition between this instability and ion-neutral damping leads to a CR streaming speed that is nearly equal to the ion Alfv\'en speed, independent of CR energy up to $\sim 1$ TeV. Consequently, CR diffusion occurs exclusively via field line random walk along magnetic field lines, and the diffusion coefficient for this process is $D\sim 10^{27}-10^{28}$ cm$^2$ s$^{-1}$ independent of CR energy below $\sim 1$ TeV. At higher energies CRs are able to stream faster than the ion Alfv\'en speed, and this leads to an increase in the effective diffusion coefficient.

We then use the diffusion coefficients we derive to analyse CR calorimetry and $\gamma$-ray production in the starburst galaxies NGC 253, M82, and Arp 220, all of which have measured $\gamma$-ray fluxes from \textit{Fermi}/LAT. We find that the energy-independent diffusion coefficients predicted by our model naturally explains why starbursts are close to being calorimetric across a wide range of energies, and that this in turn explains the relatively hard $\gamma$-ray spectra observed in starburst galaxies. Unlike in the Milky Way, where higher-energy CRs are more likely to escape and thus produce fewer $\gamma$-rays, in starbursts there is no variation in escape probability with CR energy at energies below $\sim 1$ TeV, and thus the $\gamma$-ray spectrum has approximately the same slope as the underlying CR spectrum at injection. We predict a deviation from calorimetry above $\sim 1$ TeV due to the increase in diffusion coefficient associated with faster CR streaming. We show that this prediction is consistent with existing HESS and VERITAS observations of NGC 253 and M82, and we predict a similar turn-down for Arp 220 that should be observable with the upcoming Cerenkov Telescope Array.

\section*{Acknowledgements}

The authors thank B.~Burkhart, C.~Federrath, B.~Lacki, and T.~Thompson for helpful discussions and comments on the manuscript, Jamie Holder and Olaf Reimer for comments specifically addressing the $\gamma$-ray observations of Arp 220, and an anonymous referee for helpful comments. MRK and RMC acknowledge support from the Australian Research Council through \textit{Discovery Projects} award DP190101258. MRK acknowledges support from a Humboldt Research Award from the Alexander von Humboldt Foundation. SX acknowledges the support for Program number HST-HF2-51400.001-A provided by NASA through a grant from the Space Telescope Science Institute, which is operated by the Association of Universities for Research in Astronomy, Incorporated, under NASA contract NAS5-26555.




\bibliographystyle{mnras}
\bibliography{refs} 



\begin{appendix}

\section{Turbulent damping of the streaming instability}
\label{app:turb_damping}

Alfv\'en waves generated by the streaming instability can be damped by ion-neutral dissipation, as discussed in the main text, but they can also be damped by non-linear interactions with Alfv\'en waves that are part of the turbulent cascade \citep{Yan02a, YL04, FG04, Lazarian16a}. In this appendix we show that this process is, however, unimportant in starburst environments. In assessing turbulent damping, we must consider two distinct questions: what wavelengths parallel to the local magnetic field (and thus what range of CR gyroradii) are subject to damping, and what is the damping rate in the energy range where damping operates? 
With regard to the first of these questions, for strong MHD turbulence, turbulent damping acts only on CRs with gyroradii that satisfy \citep{Lazarian16a}
\begin{equation}
    \rg \gtrsim \max\left(1,\MA\right) \left(\frac{l_{\rm min}}{L}\right)^{1/3} l_{\rm min},
\end{equation}
where $l_{\rm min}$ is the characteristic size of the smallest Alfv\'enic eddies in the direction perpendicular to the local magnetic field. CRs with smaller gyroradii generate Alfv\'en waves that are not in resonance with any of the waves in that are part of the turbulent cascade, and thus suffer no significant damping. For our damped cascade we have $l_{\rm min} = \LdampA$, and using \autoref{eq:LdampA} for $\LdampA$ and \autoref{eq:rg} for $\rg$, we find that turbulent damping only applies to CRs with energies $\ECR \gtrsim E_{\rm CR,damp}$, where
\begin{eqnarray}
E_{\rm CR,damp} & = &\frac{\pi^{5/6} e \uLA^3}{2^{2/3} L \gd^2 \chi^2 \left(\mH \muH \nH\right)^{3/2} \sin\alpha } \max\left(1,\MA^{-1/3}\right)
\nonumber \\
& = & 0.58 \frac{\uLAone^3}{L_2 \nHthree^{3/2} \chi_{-4}^2} \frac{\max\left(1,\MA^{-1/3}\right)}{\sin\alpha} \, \mbox{TeV}.
\end{eqnarray}
Evaluating this quantity for the sample starbursts in \autoref{tab:starburst_properties}, using the same fiducial parameter choices as those used to evaluate $\ECRscat$ in that table, gives $E_{\rm CR,damp} = 27\chi_{-4}^{-2}$, $27\chi_{-4}^{-2}$, and $68\chi_{-4}^{-2}$ TeV for Arp 220, M82, and NGC 253, respectively. Thus turbulent damping is only potentially important for CRs with energies in excess of tens of TeV.

With regard to the second question, the turbulent damping rate is given by \citep{Lazarian16a}
\begin{equation}
    \omega_{\rm td} = \frac{\vAt}{\sqrt{\rg L}} \MA^{3/2} \min\left(1,\MA^{1/2}\right).
\end{equation}
This can be compared to the ion-neutral damping rate $\omega_{\rm d}=\nuin/2$ (\autoref{eq:omegad}). The ratio of the two damping rates is
\begin{equation}
\frac{\omega_{\rm td}}{\omega_{\rm d}} =
\left(\frac{8 \sqrt{\pi} e \uLA^3}{\ECR L \sin\alpha}\right)^{1/2} \frac{\min\left(1,\MA^{1/2}\right)}{\gd \left(\muH \mH \nH\right)^{3/4}}.
\end{equation}
Since this ratio is a decreasing function of $\ECR$, we can put an upper limit on the potential importance of turbulent damping relative to ion-neutral damping by evaluating for CRs of energy $E_{\rm CR,damp}$. Doing so, we obtain the remarkably simple result
\begin{equation}
    \frac{\omega_{\rm td}}{\omega_{\rm d}} < \left(\frac{2048}{\pi}\right)^{1/6} \chi \min\left(1,\MA^{2/3}\right) \approx 2.9 \times 10^{-4} \chi_{-4} \min\left(1,\MA^{2/3}\right).
\end{equation}
The conclusion to be drawn from this expression is that, in a weakly-ionised medium, $\chi \ll 1$, turbulent damping is always sub-dominant compared to ion-neutral damping, by an amount of order $\chi$. We therefore neglect turbulent damping in the main text.

\section{Dependence of calorimetry on numerical parameter choices}
\label{app:diff_numerical}

In this appendix we explore how our results for the degree of calorimetry in starburst systems depend on the functional dependence of the diffusion coefficient on gas density and local CR density, $D\propto n^{-\beta} U^{-\eta}$, and on the choice of upper boundary condition.

\subsection{Dependence on gas density}

\begin{figure}
    \centering
    \includegraphics[width=\columnwidth]{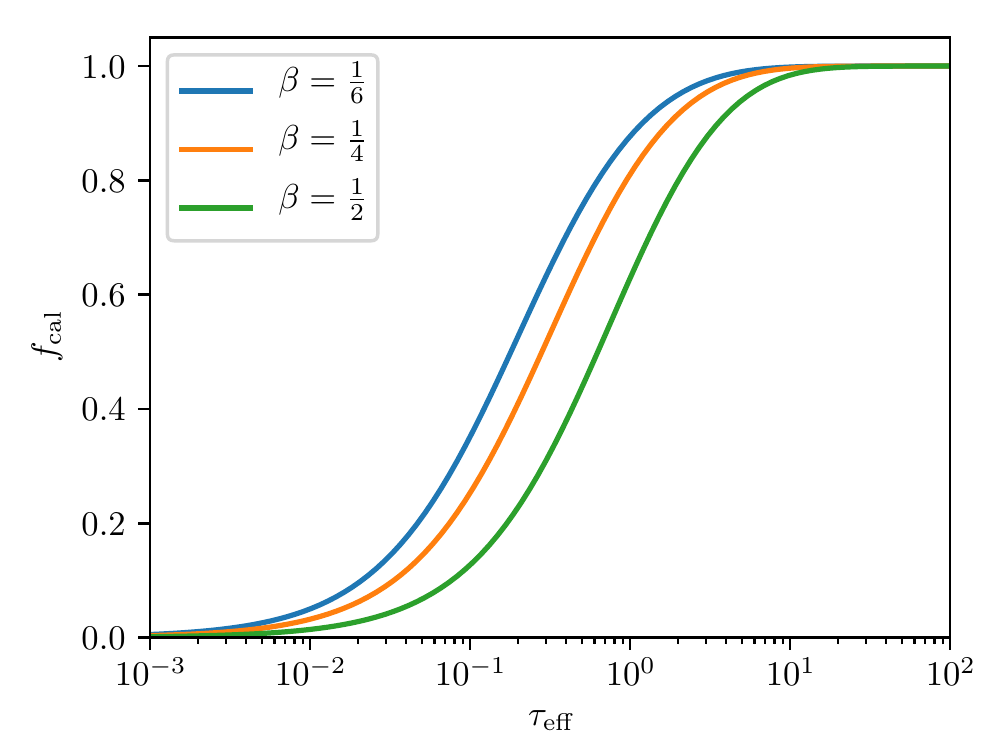}
    \caption{Calorimetric fraction $f_{\rm cal}$ as a function of effective optical depth $\tau_{\rm eff}$ for our fiducial scaling of diffusion coefficient with density ($\beta=1/4$), and for two alternative models where the diffusion coefficient is either significantly less ($\beta=1/6$) or significantly more ($\beta=1/2$) density-dependent. The plotted lines use $\Upsilon_0=10.67$, but the results are nearly indistinguishable with any $\Upsilon_0\gtrsim 1$.}
    \label{fig:diff_beta}
\end{figure}

We first explore how the degree of calorimetry changes in response to $\beta$. Plausible values of $\beta$ range from $\beta\approx 1/6$ (expected for Kolmogorov turbulence in an ionised medium, or for streaming in a medium where the ionisation fraction is varies more weakly with density that $\chi\propto n^{1/2}$), to $\beta\approx 1/2$ expected for a medium where $\chi \propto n$. We show $f_{\rm cal}$ versus $\tau_{\rm eff}$ for these alternatives in \autoref{fig:diff_beta}. Clearly the scaling of diffusion coefficient with density has relatively minor effects on calorimetry. For our fiducial value $\beta=1/4$, galaxies become 50\% calorimetric at $\tau_{\rm eff}\approx 0.28$ (for our fiducial $\Upsilon_0=10.67$, though changing $\Upsilon_0$ has only very minor effects), whereas changing $\beta$ over the range $1/6 - 1/2$ causes this value to shift over the range $0.18-0.63$, a fairly minor effect.

\subsection{Dependence on CR energy density}

We next explore how calorimetry changes if $D$ depends on the local CR energy density, $\eta\neq 0$. We focus on the case $\eta=1/4$, as expected if $D\propto \vAi \propto \zeta^{-1/4}$, where $\zeta$ is the primary CR ionisation rate, and $\zeta\propto U$. In this case \autoref{eq:diffeq_nondim} becomes
\begin{equation}
    \frac{d}{dx}\left(u^{-\eta} e^{\beta x} \frac{du}{dx}\right) = \tau_{\rm eff} e^{-x} u.
    \label{eq:diffeq_num}
\end{equation}
The boundary condition at $x=0$ is still $u(0)=1$, and the boundary condition at $x\to\infty$ becomes
\begin{equation}
    -\left(u^{-\eta} e^{\beta x}\frac{d\ln u}{dx}\right)_{x\to\infty} = \Upsilon_0.
    \label{eq:bc_num}
\end{equation}

We cannot solve \autoref{eq:diffeq_num} analytically, but we can obtain a solution numerically. Since \autoref{eq:diffeq_num} is a boundary value problem, we use a shooting method. Starting at $x=x_{\rm max}$ for $x_{\rm max}\gg 1$ so that the flux is near its asymptotic value (we use $x_{\rm max}=30$ for the numerical results below, which we have confirmed is sufficient for converged results), we guess a value of $u'(x_{\rm max})$, and then use \autoref{eq:bc_num} to obtain the corresponding value of $u(x_{\rm max})$. We then integrate backward to $x=0$, and evaluate $u(0)$. We do this repeatedly, using Newton's method to search for search for a value of $u'(x_{\rm max})$ such that $u(0)=1$, our other boundary condition. The result is a numerical solution to \autoref{eq:diffeq_num} that satisfied both boundary conditions. We can then use this solution to evaluate the calorimetric fraction
\begin{equation}
    f_{\rm cal} = 1 - \frac{\left(u^{-\eta} e^{\beta x} u'\right)_{x=x_{\rm max}}}{\left(u^{-\eta} e^{\beta x} u'\right)_{x=0}} = 1 - \frac{\left(u^{-\eta} e^{\beta x} u'\right)_{x=x_{\rm max}}}{\left(u'\right)_{x=0}}.
    \label{eq:fcal_num}
\end{equation}

\begin{figure}
    \centering
    \includegraphics[width=\columnwidth]{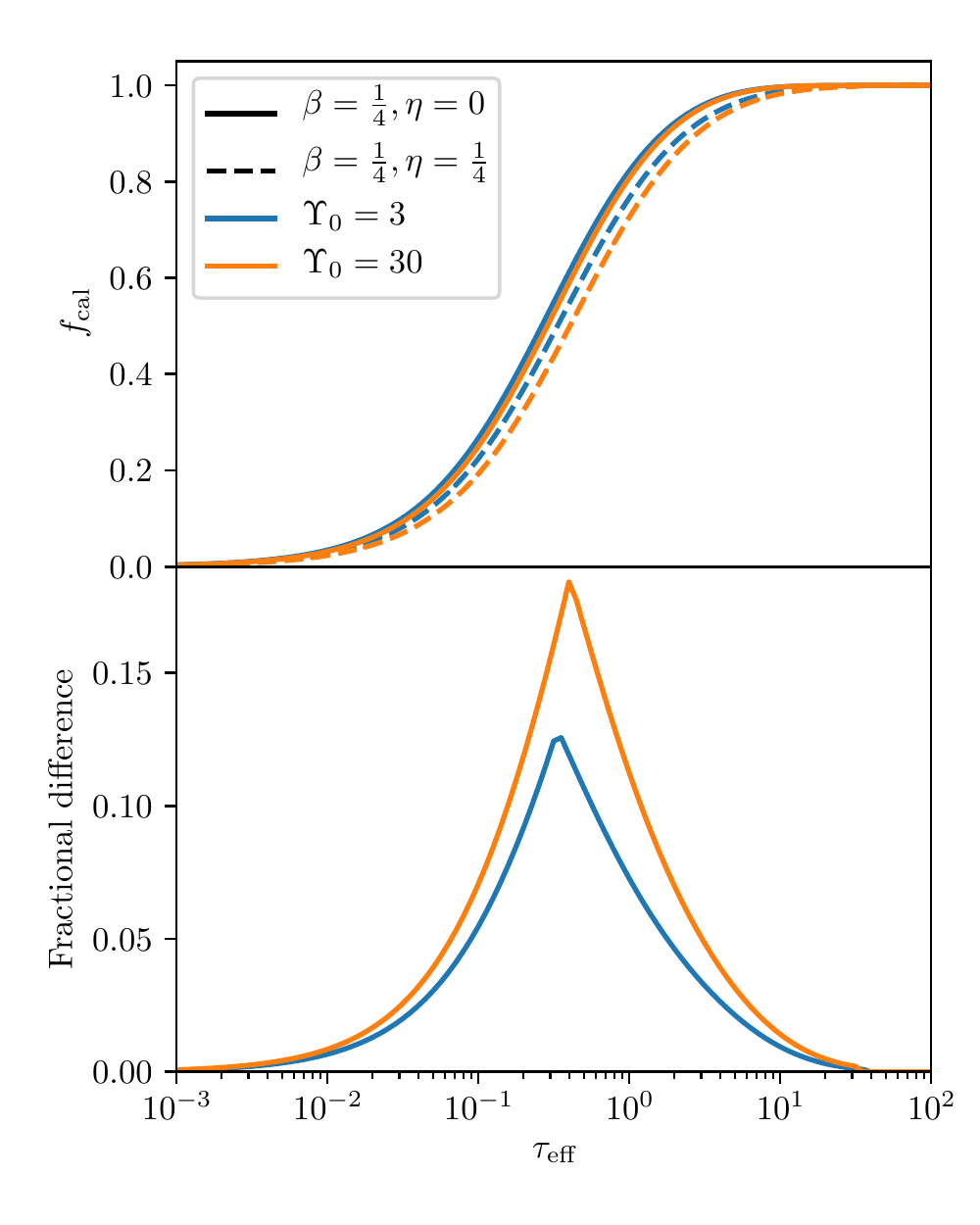}
    \caption{Comparison of the calometric fractions $f_{\rm cal}$ derived from the analytic result for $\beta=1/4, \eta=0$ (\autoref{eq:fcal}) versus using \autoref{eq:fcal_num} together with the numerical solution to \autoref{eq:diffeq_num} for $\beta=1/4,\eta=1/4$. The top panel shows $f_{\rm cal}$ as a function of $\tau_{\rm eff}$ computed both ways, for the cases $\Upsilon_0=3$ and $30$. The bottom panel shows the fractional difference between the two estimates, defined as $(f_{\rm cal,\eta=0}-f_{\rm cal,\eta=1/4}) / \max(f_{\rm cal,\eta=1/4}, 1-f_{\rm cal,\eta=1/4})$.}
    \label{fig:diff_numerical}
\end{figure}

We show $f_{\rm cal}$ as a function of $\tau_{\rm eff}$ as determined via this numerical procedure for $\beta=1/4$, $\eta=1/4$, and compare to the analytic solution given by \autoref{eq:fcal} for the case $\beta=1/4$, $\eta=0$, in \autoref{fig:diff_numerical}. The figure shows that the two solutions differ by at most $\approx 15\%$. 

\subsection{Dependence on the outer boundary condition}

Our fiducial choice of boundary condition is to require that the CR flux approach the free-streaming flux as $z\to\infty$. However, this is obviously a simplification. While the midplanes of starbursts are dominated by cold, molecular gas that is not being ejected in bulk, the regions $\gtrsim 10-20$ scale heights above the midplane are usually dominated by an ionised outflow \citep[e.g.,][]{Veilleux05a} into which the CRs will be locked \citep[e.g.][]{Breitschwerdt91a, Dorfi12a}. One might therefore imagine as an alternate boundary condition that we should join our diffusion solution for the region near the midplane onto a wind advection solution for the region above it.

In terms of our simple diffusion system, \autoref{eq:diffeq}, this amounts to requiring that the effective CR advection velocity associated with the diffusive flux approach the flow speed at the base of the wind, i.e., $-D(dU/dz) \to (4/3) V_{\rm w} U$, where $V_{\rm w}$ is the speed at the base of the wind, where the CRs become locked with it. This amounts to replacing $\vst$ with $V_{\rm w}$ in \autoref{eq:upsilon},i.e.,
\begin{equation}
    \Upsilon_0 = \frac{4 h V_{\rm w}}{3 D_0}
    = 41.2 h_2 V_{\rm w,2} D_{0,27}^{-1} = \frac{4}{3} \MA^3 \frac{V_{\rm w}}{\vst},
    \label{eq:upsilon_wind}
\end{equation}
where $V_{\rm w,3} = V_{\rm w}/10^3$ km s$^{-1}$.

\begin{figure}
    \centering
    \includegraphics[width=\columnwidth]{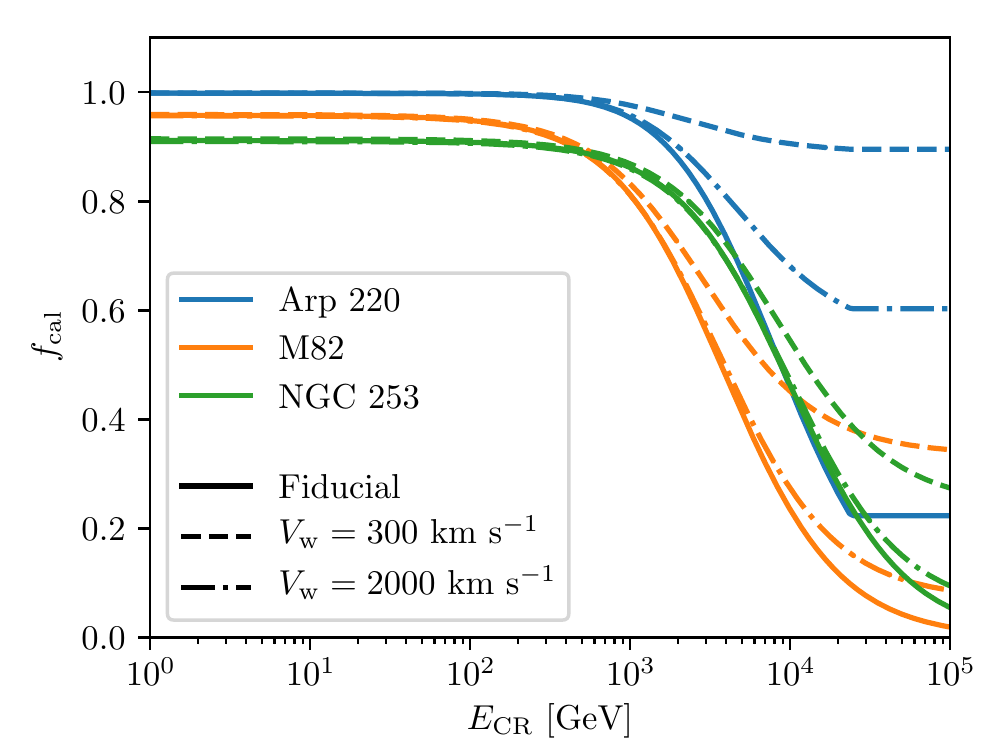}
    \caption{Calorimetric fraction $f_{\rm cal}$ for our fiducial parameters for Arp 220 (blue), M82 (orange), and NGC 253 (green), and three different treatments of the upper boundary condition. The solid lines show our fiducial choice (\autoref{eq:upsilon}), in which we assume that the CR flux approaches the streaming flux as $z\to\infty$. The dashed and dot-dashed lines correspond to \autoref{eq:upsilon_wind}, where we assume the flux approaches a value corresponding to a constant advection speed at a wind velocity of $V_{\rm w} = 300$ km s$^{-1}$ (dashed lines) or $1000$ km s$^{-1}$ (dot-dashed lines), which span the plausible range of wind velocities.
    }
    \label{fig:bc}
\end{figure}

In \autoref{fig:bc} we show calculations of the predicted calorimetric for Arp 220, M82, and NGC 253 using our fiducial parameters for these galaxies (\autoref{tab:starburst_properties}) using both our fiducial boundary condition where we assume that CRs approach the streaming molecular gas streaming speed as $z\to\infty$ (\autoref{eq:upsilon}) and the alternative boundary condition given by \autoref{eq:upsilon_wind}, computed using $V_{\rm wind} = 300$ and 2000 km s$^{-1}$. These represent values toward the very low end and the very high end of observationally-inferred velocities for the hot components of starburst winds \citep{Veilleux05a}. Thus, if the alternative wind boundary condition is preferable, the range of plausible solutions lies between these two lines.

We see that using the alternative boundary condition has completely negligible effects on our inferred calorimetric fractions for any galaxy at energies $\lesssim 1$ TeV, and that is has only minor effects for M82 or NGC 253 up to 100 TeV. The only galaxy and energy range for which there is a substantial effect is Arp 220 at energies $\gtrsim 1$ TeV, where the change in boundary condition plausibly moves the system from mostly non-calorimetric to mostly calorimetric. The effect here is significant because, due to the extraordinarily strong ion-neutral damping in this galaxy, the CR streaming speed in the ionised medium approaches $c$ for energies $E_{\rm CR}\gtrsim 10$ TeV. If we impose as a boundary condition that the CR effective speed cannot exceed $V_{\rm w}=300$ km s$^{-1}$ in the wind above the midplane, then high-energy CRs have far more difficulty escaping, and the calorimetric fraction rises correspondingly. Future observations of Arp 220 in $\gtrsim 1$ TeV gamma-rays should be able to determine if the escape of $\gtrsim 10$ TeV CRs is in indeed choked off by an ionised wind layer that traps the CRs that cannot be confined by the molecular medium. At present, however, observations provide only upper limits on the TeV $\gamma$-ray flux, and these are not sufficiently sensitive to provide useful constraints (c.f. \autoref{fig:grspectra}).

\end{appendix}


\bsp	
\label{lastpage}
\end{document}